\documentclass[12pt,a4paper]{article}                           

\usepackage{a4}          

\usepackage{amsmath}
\usepackage{amssymb}
\usepackage{graphicx}
\usepackage[left=3.2cm, right=3.2cm, top=4cm, bottom=3cm]{geometry}
\usepackage[compact]{titlesec}
\usepackage[ansinew]{inputenc}
\usepackage{html}
\usepackage{appendix}
\usepackage{fancyhdr}
\usepackage{german}
\usepackage{cite}
\usepackage[english]{babel}
\usepackage{sectsty}

\def\hybrid{\topmargin -20pt    \oddsidemargin 0pt
        \headheight 0pt \headsep 0pt
        \textwidth 6.25in       
        \textheight 9 in       
        \marginparwidth .875in
        \parskip 5pt plus 1pt 
          \jot = 1.5ex
   }
\hybrid
\numberwithin{equation}{section}
\numberwithin{table}{section}

\newcommand{\dd}{\mathrm{d}}
\newcommand{\BB}{\mathcal{B}}
\newcommand{\Tr}{\mathrm{Tr}}
\newcommand{\MM}{\mathcal{M}_{\mathcal{N}=1}}
\newcommand{\DD}{\mathcal{D}}
\newcommand{\NN}{\mathcal{N}}
\newcommand{\nn}{\nonumber}

 \newcommand{\bea}{\begin{eqnarray}}
 \newcommand{\eea}{\end{eqnarray}}

\titlespacing{\section}{0pt}{25pt}{15pt}
\titlespacing{\subsection}{0pt}{20pt}{10pt}

\begin{document}
\baselineskip=14pt
\parskip 5pt plus 1pt

\vspace*{-1.5cm}
\begin{flushright}    
  {\small
 
 }
\end{flushright}

\vspace{2cm}
\begin{center}        
  {\LARGE
  The effective action of D6-branes in $\mathcal{N} =1$ type IIA orientifolds  
  }
\end{center}

\vspace{0.75cm}
\begin{center}        
Max Kerstan, Timo Weigand
\end{center}

\vspace{0.15cm}
\begin{center}        
  \emph{ Institut f\"ur Theoretische Physik, Ruprecht-Karls-Universit\"at Heidelberg,\\
             Philosophenweg 19, 69120 Heidelberg, Germany }
\end{center}

\vspace{2cm}


\begin{abstract}
We use a Kaluza-Klein reduction to compute the low-energy effective action for the massless modes of a spacetime-filling D6-brane wrapped on a special Lagrangian 3-cycle of a type IIA Calabi-Yau orientifold. The modifications to the characteristic data of the $\mathcal{N}=1$ bulk orientifold theory in the presence of a D6-brane are analysed by studying the underlying Type IIA supergravity coupled to the brane worldvolume in the democratic formulation and performing a detailed dualisation procedure. The ${\cal N}=1$ chiral coordinates are found to be in agreement with expectations from mirror symmetry. We work out the K\"ahler potential for the chiral superfields as well as the gauge kinetic functions for the bulk and the brane gauge multiplets including the kinetic mixing between the two. The scalar potential resulting from the dualisation procedure can be formally interpreted in terms of a superpotential. Finally, the gauging of the  Peccei-Quinn shift symmetries of the complex structure multiplets reproduces the D-term potential enforcing the calibration condition for special Lagrangian 3-cycles.

\end{abstract}

\thispagestyle{empty}

\clearpage

\tableofcontents

\setcounter{page}{1}
\setlength{\headsep}{30pt}

\section{Introduction}
A large class of phenomenologically attractive string theory vacua has been constructed by compactifying type II string theory with D-branes on Calabi-Yau manifolds or orientifolds thereof. Intersecting D-branes are particularly appealing for model building purposes because the open strings ending on them lead to a low-energy effective theory with a gauge group in which the Standard Model gauge group can be easily embedded. Hidden sector D-branes separated from the Standard Model stacks and instantonic D-branes can induce potentials suitable to stabilize moduli fields. A summary of the developments and literature centered around spacetime-filling and Euclidean branes can be found e.g. in  \cite{LuestIntBranes, blumenhagenCvetic,blumenhagen4Dcomp,DbraneModelMarchesano,Blumenhagen:2009qh,Cvetic:2011vz}.

Models of the above type consist of stacks of D-branes placed in a background which is a product of Minkowski space with a compact Calabi-Yau orientifold. Closed strings propagating in the bulk space give rise to a gravitational multiplet as well as a set of abelian vector multiplets and chiral moduli multiplets. The open strings ending on stacks of D-branes lead to non-abelian vector multiplets and charged chiral matter. For non-rigid D-branes additional uncharged moduli parametrizing the brane deformations and Wilson lines of the gauge field appear. 

The phenomenologically most promising models involve D-branes respecting the $\mathcal{N}=1$ supersymmetry of the orientifold bulk. In a realistic theory this supersymmetry, which is unbroken at the compactification scale, must be spontaneously broken at a lower energy scale. The specific form of the F- and D-term potentials inducing this spontaneous SUSY breaking must be determined from a reduction of the action. Apart from their relevance for supersymmetry breaking such potentials are also necessary to stabilize the many massless moduli fields arising in a typical string theory compactification. 
One way to obtain such potentials in the effective theory is to allow for non-vanishing background fluxes of the bulk fields. Supersymmetry breaking in the bulk sector is then transferred to the visible sector through couplings of the bulk and brane fields. Such flux compactifications have received widespread interest in the last few years, and the many developments in this field are reviewed for example in \cite{GranaFlux, DouglasKachruFlux}.

In order to determine the precise form of the soft supersymmetry breaking terms one needs to compute the low-energy action of D-branes encoding these couplings. This task has been carried out for D3-, D5- and D7-branes in type IIB orientifolds in \cite{SoftSusyBreakingGranaGrimm,Jockers:2008pe,grimmD5, jockersD7,Jockers:2005zy}. However, to our knowledge a complete reduction of of the D6-brane action has not yet been carried out. This note aims to fill this gap in the literature. For simplicity we will focus on a single brane, although many of the results can be carried over without problems to the non-abelian case. Rather than consider an explicit model we will compute the effective four-dimensional theory in terms of the intersection numbers and geometric data of a general orientifold. This will be accomplished via a Kaluza-Klein reduction of the Dirac-Born-Infeld (DBI) and Chern-Simons (CS) brane actions describing the brane dynamics at leading order in $\alpha'$ and the string coupling. We will only compute the bosonic part of the action; this suffices to completely determine also the fermionic part due to $\mathcal{N}=1$ supersymmetry.

The superpotential induced by introducing closed string fluxes in a type IIA orientifold compactification has already been evaluated in ref. \cite{grimmOrientifold}. As the D-branes couple directly to the potentials rather than the field strengths of the closed string fields this computation is not modified by the addition of a D6-brane (see \cite{SoftSusyBreakingGranaGrimm} for an example of this). For this reason we will not consider turning on bulk background fluxes in this paper. However, we will include spacetime 3-forms in the Kaluza-Klein reduction of the string fields. These fields are non-dynamical in the four-dimensional effective theory and are dual to constants. As we will see, upon dualization there appears a superpotential which forces these additional parameters to be zero (see also \cite{grimmD5, grimmOrientifold, louisMicu}).

The outline of the paper is as follows. In section \ref{IIAorientifolds} we briefly review the compactification of type IIA string theory on a Calabi-Yau orientifold. We mainly follow \cite{grimmOrientifold}, but we use the democratic formulation \cite{bergshoeffIIAsugra} of the type IIA supergravity.  In this formulation one includes in addition to the usual Ramond-Ramond p-form fields $C^{(1)}$, $C^{(3)}$ their magnetic duals $C^{(5)}$ and $C^{(7)}$ and eliminates the additional degrees of freedom by imposing suitable duality relations at the level of the equations of motion. This formulation facilitates the inclusion of D-branes into the theory as their magnetic couplings to the RR forms can be written explicitly in terms of $C^{(5)}$ and $C^{(7)}$.

The contribution of the D6-brane is discussed in section \ref{brane}. In sections \ref{susyCond} and \ref{braneModuli} we discuss the geometric conditions imposed by supersymmetry  upon  embedding the brane  as well as  their implications for the massless spectrum of the brane theory. These supersymmetry conditions depend on the bulk moduli so that the bulk and brane moduli spaces are intrinsically intertwined. The combined open-closed moduli space is best described using the concept of relative cohomology \cite{LercheMayr_SpecGeomOpenClosed}. However, this abstract formulation makes it difficult to explicitly perform the dimensional reduction of the action. We therefore work out the action in the limit of small fluctuations around a chosen background value of the moduli. In this approximation the moduli space can be treated as a product of separate brane and bulk moduli spaces. In sections \ref{DBIsection} and \ref{CSreduction} we perform the dimensional reduction of the DBI and CS actions.

In section \ref{4daction} we combine the bulk and brane contributions to the four-dimensional action and recast this action in the canonical form of an $\mathcal{N}=1$ supergravity. The dualization procedure to remove the redundant degrees of freedom introduced in the democratic formulation is performed in section \ref{dualization}. In section \ref{gaugeKin} we discuss the gauge kinetic function for the bulk and the brane gauge multiplets. This allows us to determine the  $\mathcal{N}=1$ chiral coordinates. In particular, we find that the open string moduli correct the definition of the  superfields associated with the complex structure moduli, but not the ones associated with the K\"ahler moduli, in agreement with expectations from mirror symmetry. Furthermore we find that generically the brane and the bulk gauge fields undergo kinetic mixing in the sense that the gauge kinetic function is not block-diagonal. In section \ref{kaehlerPot} we work out the K\"ahler potential in the abovementioned approximation that the bulk and the brane moduli spaces form a direct product as appropriate for small fluctuations around a supersymmetric configuration. The scalar potential arising from the dualisation procedure can formally be understood, as observed in section  \ref{superpot},  in terms of a superpotential which we discuss. Consistently the gauging of the Peccei-Quinn symmetry enjoyed by the complex structure moduli reproduces the well-known D-term potential incorporating the special Lagrangian calibration conditions.
A summary of our main findings is given in section \ref{conclusion}.

\section{Type IIA on Calabi-Yau Orientifolds}
\label{IIAorientifolds}
We begin this paper by giving a brief review of the compactification of type IIA string theory on a Calabi-Yau orientifold.  In section \ref{Ospectrum} we determine the massless 4-dimensional spectrum of type IIA supergravity in the democratic formulation of \cite{bergshoeffIIAsugra}. In section \ref{section_Bulk} we integrate out the internal space to obtain the effective 4-dimensional theory. The reduction of the curvature scalar is carried out in detail in appendix \ref{reduxR}.

\subsection{The massless spectrum}
\label{Ospectrum}
\noindent The setup we are considering involves a D6-Brane wrapped on a 3-cycle of a type IIA Calabi-Yau orientifold. The $\mathcal{N}=2$ supersymmetric spectrum of  type IIA string theory is modded out to $\mathcal{N}=1$ by requiring $(-1)^{F_L}\Omega_p \sigma$ to be a symmetry, as detailed in\cite{grimmOrientifold}. Here $F_L$ is the left-moving spacetime fermion number, $\Omega_p$ is the worldsheet parity while $\sigma$ is an involution of the Calabi-Yau manifold $Y$ which acts by pullback on the fields of our effective theory.\\
One can show that in order to preserve $\mathcal{N}=1$ supersymmetry the involution $\sigma$ must be antiholomorphic and isometric \cite{orientifoldMirrorSymm_Vafa}, i.e. the K\"ahler form of the Calabi-Yau three-fold $Y$ must transform according to
\begin{equation}\sigma^* J = -J.\label{pullbackJ}\end{equation}
Due to the antiholomorphicity of $\sigma$ the pullback of the holomorphic (3,0)-form $\Omega$ of $Y$ must be some multiple of $\overline{\Omega}$, and the compatibility of the pullback with the Calabi-Yau condition $\Omega\wedge\overline{\Omega}\propto J\wedge J\wedge J$ implies
\begin{equation}\exists \, \theta\!\in[0,2\pi):\qquad\sigma^* \Omega = e^{2i\theta}\,\overline{\Omega}.\label{pullbackOmega}\end{equation}
Note that the holomorphic (3,0)-form $\Omega$ on a Calabi-Yau manifold is not uniquely defined by the properties listed above. Instead, one has the freedom to rescale by $e^{h(z)}$ with an arbitrary holomorphic function $h$ on the complex structure moduli space. The phase $\theta$ is determined by the choice of phase for $\Omega$, as a rescaling $\Omega\rightarrow e^{i\varphi}\Omega$ changes $\theta\rightarrow \theta+\varphi$. However, once one has fixed a particular value of $\theta$ in this manner this rescaling freedom is reduced to a real rescaling by $e^{\mathrm{Re}(h(z))}$ on an $\mathcal{N}=1$ orientifold by the condition (\ref{pullbackOmega}).\\
\indent At the fixed point locus of $\sigma$ we include an O6 orientifold plane to cancel the RR-charge and tension of the D6-branes in order to eliminate RR and gravitational anomalies. The above conditions imply that the cycle wrapped by this O6-plane is a special Lagrangian cycle calibrated with respect to the form $\mathrm{Re}(e^{i\theta}\Omega)$. As the orientifold planes do not introduce any additional degrees of freedom and their action is identical to that of a D6-brane after setting the world-volume fields to zero we do not need to treat this object separately.
\\
\indent The massless bosonic part of the closed string spectrum of type IIA supergravity contains the metric, dilaton and antisymmetric two-form $B^{(2)}$ from the NS-NS sector of the theory, as well as the p-form fields $C^{(p)},\; p=1,3$ from the RR-sector. In the democratic formulation we will use additionally the dual form fields $C^{(5)},$ $C^{(7)}$, where the additional degrees of freedom introduced in this way are eliminated by imposing a set of duality relations at the level of the equations of motion. This will be helpful as the Chern-Simons action is best expressed using all the forms $C^{(p)},\; p=1,3,5,7$. The duality relations to be imposed are \cite{bergshoeffIIAsugra, bandosDemocSugra}
\begin{eqnarray}
 G^{(6)} &=-\ast_{10} G^{(4)}, & G^{(8)}=\ast_{10}G^{(2)}\; ,\nn\\
 G^{(2)}& :=\mathrm{d}C^{(1)} ,& G^{(p)}:=\mathrm{d}C^{(p-1)}-\mathrm{d}B^{(2)}\wedge C^{(p-3)}, \label{dualityG}\end{eqnarray}
modulo forms that are exact on $Y$ and do not contribute to the 4d action.
\\
\indent We take a block-diagonal ansatz for the metric
\begin{equation}\hat{g}=\eta_{\mu\nu}\dd x^{\mu}\otimes\dd x^{\nu}+ g_{mn}\dd y^m\otimes \dd y^n. \label{metric}\end{equation}
Hence our Laplacian splits as $\Delta_{10}=\Delta_4+\Delta_Y$ and the massless 4d fields correspond to harmonic forms on $Y$ so we expand our fields into the different nontrivial cohomology groups $H^{(p,q)}(Y)$ on $Y$.
\\
\indent Note that in the presence of objects with non-vanishing tension such as D-branes and orientifold planes the ansatz taken in (\ref{metric}) is actually not a solution of the Einstein equations. In a consistent compactification one should rather take a warped metric ansatz of the form $\hat{g}=\Delta^{-1}(y)\eta+\Delta(y)g$. Equation (\ref{metric}) is actually to be read as an approximation to this warped metric, which is valid in the limit of large radius where the warp factor varies slowly over the internal space.\\
\indent As $\mathcal{N}=2$ supersymmetry requires the compactification manifold to remain a Calabi-Yau space only such variataions of the internal metric are allowed which respect this condition. In particular, the variations $\delta g_{ij}$ are restricted by the requirement that the resulting manifold remains Ricci-flat. This leads to the Lichnerowicz equation for the deformations, and as is well-known the solutions of this equation split into two classes, the deformations of the K\"ahler structure and the complex structure \cite{moduliCandelas}.\\
The requirement of Ricci-flatness implies that the variations of the mixed holomorphic and antiholomorphic components of the string frame metric form the components of a harmonic form. Hence they can be identified as the variations of the K\"ahler structure of the manifold and may be expanded into a basis $\left\{\omega_A\right\}$ of $H^{(1,1)}(Y)$ as 
\begin{equation}\delta g_{i\bar{j}}= - i \delta J_{i\bar{j}}=-i v^A(\omega_A)_{i\bar{j}}. \label{var_g_mixed}\end{equation}
The variations of the purely holomorphic or purely antiholomorphic components of the string frame metric in such a Calabi-Yau compactification are described by a set of $h^{(2,1)}$ complex scalar moduli fields $z^K$ and can be related to a basis of (2,1)-forms $\chi_K$ via 
\begin{equation}\delta g_{\bar{i}\bar{j}}=\frac{-1}{||\Omega ||^2}\bar{\Omega}_{\bar{i}}^{\;kl}(\chi_{K})_{kl\bar{j}}z^K.\label{var_g_pure}\end{equation}
The Lichnerowicz equation is exactly the requirement that the set of functions $(\chi_{K})_{kl\bar{j}}$ are the coefficients of a harmonic differential form. As the deformed manifold should still be Calabi-Yau and in particular K\"ahler there must exist a complex coordinate system in which these pure components of the deformed metric vanish. In this sense these variations can be identified with a variation of the complex structure on the underlying real manifold. This in turn can be identified with the variation of the (3,0) form $\Omega$. The forms $\chi_K$ are related to the variation of $\Omega$ via Kodaira's formula 
\begin{equation}\chi_K(z)=\partial_{z^K}\Omega(z)+\Omega(z)\partial_{z^K}K_{CS},  \label{Kodaira}  \end{equation}
where $K_{CS}$ is the K\"ahler potential defined in (\ref{CSmetric}).\\
\indent To implement the orientifold projection including the action of the pullback of $\sigma$ it is helpful to split the cohomology spaces further into eigenspaces of $\sigma$. As $\sigma$ is antiholomorphic, the spaces that are left invariant are $H^{(1,1)}, H^{(2,2)}$ and $H^3$, and we can split these into spaces consisting of forms with even or odd parity under pullback along $\sigma$ \cite{grimmOrientifold}. Finally we note that the volume form on $Y$ is proportional to $J\wedge J\wedge J$ and is therefore odd under $\sigma^*$. As the Hodge star involves contraction with the volume form, we see that it maps even forms to odd ones and vice versa, i.e. it induces an isomorphism from $H_+^{(1,1)}$ to $H_-^{(2,2)}$ etc.. In summary, following the notation of \cite{grimmOrientifold} the forms used in the Kaluza-Klein reduction of our fields are given in table \ref{basisforms}.
\begin{table}[ht]
	\centering
		\begin{tabular}{|l||c|c|c|c|c|c|}
		\hline cohomology group & $H^{(1,1)}_+$ & $H^{(1,1)}_-$ & $H^{(2,2)}_+$ & $H^{(2,2)}_-$ & $H^3_+$ & $H^3_-$\\ \hline dimension & $h_+^{(1,1)}$ & $h_-^{(1,1)}$ & $h_-^{(1,1)}$ & $h_+^{(1,1)}$ & $h^{(2,1)} + 1$ & $h^{(2,1)} + 1$ \\ \hline
		real basis  & $\omega_{\alpha}$ & $\omega_a$ & $\tilde{\omega}^a$ & $\tilde{\omega}^{\alpha}$ & $\alpha_{\hat{K}}$ & $\beta^{\hat{K}}$ \\ \hline
		\end{tabular}
		\caption{Cohomology groups and their bases.}
		\label{basisforms}
\end{table}
\\The basis elements are chosen to satisfy
\begin{equation}
\begin{array}{rclcl}\int_Y \omega_{\alpha}\wedge\tilde{\omega}^{\beta} &=& l_s^6\delta_{\alpha}^{\beta}, &\quad& \alpha, \beta \in \{1,\ldots,h^{(1,1)}_+\}, \\
 \int_Y \omega_a\wedge\tilde{\omega}^b &=& l_s^6\delta_a^b, &\quad& a, b \in \{1,\ldots,h^{(1,1)}_-\}, \\
\int_Y \alpha_{\hat{K}}\wedge \beta^{\hat{L}} &=& l_s^6\delta_{\hat{K}}^{\hat{L}}, &\quad& \hat{K}, \hat{L} \in \{1,\ldots,h^{(2,1)}+1\},
\label{baseNormalization}
\end{array}
\end{equation}
where we introduced the string length
\bea
l_s = 2 \pi \sqrt{\alpha'}.
\eea
\indent We are now ready to perform the Kaluza-Klein reduction to obtain the massless effective 4-dimensional fields. To do this, we note that under $(-1)^{F_L}\Omega_p$ the forms $C^{(p)},\;p=3,7$ as well as the metric and dilaton are even, while $B^{(2)}$ and $C^{(p)},\;p=1,5$ are odd. The resulting expansions of the fields read
\begin{eqnarray}
C^{(1)}&=&0,  \quad\quad
C^{(3)}=  c_{(3)}+V^{\alpha}\wedge\omega_{\alpha} + \xi^{\hat{K}}\alpha_{\hat{K}}, \nn\\
C^{(5)}&=&\tilde{c}_{(3)}^a\wedge\omega_a+U_{\alpha}\wedge\tilde{\omega}^{\alpha}+\rho_{(2)\hat{K}}\wedge \beta^{\hat{K}},  \quad\quad
C^{(7)}=d_{(3)a}\wedge\tilde{\omega}^a,\nn\\
B^{(2)}&=&b^a \omega_a, \quad\quad
J=v^a \omega_a,
\label{fieldExp}
\end{eqnarray}
where in the last assertion the property \ref{pullbackJ} for $J$ was used.
Note that we have included three-forms on $\textbf{R}^{1,3}$ although they do not represent dynamical degrees of freedom as they are dual to constants which appear in the scalar potential of the resulting action (see e.g. refs. \cite{grimmOrientifold, louisMicu}). 
Finally, in addition to the odd B-field moduli $b^a$, the periodic shift symmetry of $B^{(2)}$ implies that one can consider a discrete B-field along elements of $H^{1,1}_+(Y)$
\bea
\label{B+}
B^{(2)}_+ = b^\alpha \omega_\alpha ,\quad\quad b^\alpha = 0 \quad {\rm or} \quad \frac12.
\eea
This parallels the situation in Type IIB orientifolds.

\indent We still have to take into account the effect of the orientifold projection on the complex structure moduli space following ref. \cite{grimmOrientifold}. Expanding $\Omega$ into the basis of three-forms above,
\begin{equation}\Omega(z)=\mathcal{Z}^{\hat{K}}\alpha_{\hat{K}}-\mathcal{F}_{\hat{K}}\beta^{\hat{K}},\end{equation}
and inserting it into (\ref{pullbackOmega}) we obtain the conditions\footnote{In fact, we can be more general. E.g. by mirror symmetry with Type IIB theory, which maps the Type IIB K\"ahler moduli to the periods $U_K \equiv {\cal Z}_K/ {\cal Z}_0$, it is clear that the latter enjoy a shift symmetry. Just as the shift symmetry of the K\"ahler moduli allows for the discrete $B^{(2)}_+$-field as in (\ref{B+}) in agreement with the orientifold involution, there is a ${\mathbb Z}_2$ freedom ${\rm Im}( U_K) =0$ or $1/2$ (for $\theta =0$). See \cite{Bachas:2008jv} for more details.} 
\begin{eqnarray}{\rm Im}(e^{-i\theta}\mathcal{Z}^{\hat{K}})=0, \quad\quad\quad  {\rm Re}(e^{-i\theta}\mathcal{F}_{\hat{K}})=0.\label{condF}\end{eqnarray}
The two conditions are not independent because via Kodaira's formula one set of periods (e.g. the $\mathcal{F}_{\hat{K}}$) can be shown  to be a function of the other ones. In fact the $\mathcal{F}_{\hat{K}}$ are given by the derivatives of a holomorphic prepotential which is a function of the $\mathcal{Z}^{\hat{K}}$ \cite{moduliCandelas}. In other words, the moduli space of complex structure deformations of a Calabi-Yau space is a special K\"ahler manifold. Hence we take the first condition in (\ref{condF}) as a constraint on the allowed $z^K$, and the second condition in (\ref{condF}) is to be read as a constraint on the periods of $Y$.  This is a constraint on the prepotential which must be fulfilled for $Y$ to admit an antiholomorphic involution satisfying (\ref{pullbackOmega}).\\
\indent We may use the phase of the complex rescaling freedom to set one of the ${\rm Im}(e^{-i\theta}\mathcal{Z}^{\hat{K}})$ to zero. Then the first constraint in (\ref{condF}) gives us $h^{(2,1)}$ real equations allowing us to eliminate exactly half of the degrees of freedom of the original $h^{(2,1)}$ complex fields $z^K$. With the help of Kodaira's formula it is possible to show that the surviving degrees of freedom can be identified with the real or imaginary parts of the original complex fields (which were defined by (\ref{var_g_pure})), and that the $\mathcal{N}=1$ complex structure moduli space spans a Lagrangian submanifold of the corresponding $\mathcal{N}=2$ moduli space \cite{grimmOrientifold}. Local coordinates for $\mathcal{M}^{CS}_{\mathcal{N}=1}$ are given by a set of real fields $q^K$ such that for each $K\in \left\{1,\ldots, h^{(2,1)}\right\}$ there holds either
\bea
q^K=z^K \quad\quad {\rm or} \quad\quad  iq^K=z^K.
\eea

\subsection{The effective bulk action}
\newcommand{\KK}{\mathcal{K}}
\label{section_Bulk}
\noindent We are now ready to determine the effective 4d bulk action by inserting the expansions of the fields from (\ref{fieldExp}) into the democratic form of the (bosonic part of the) IIA supergravity pseudo-action in the string frame as given in refs. \cite{bergshoeffIIAsugra, bandosDemocSugra},
\begin{eqnarray}
S^{(10)}_{IIA}&=&\frac{1}{2\kappa_{10}^2}\int_{\mathbf{R}^{1,3}\times Y}e^{-2\phi}(\dd^{10} x\sqrt{-\det \hat{g}}\;(-\hat{R})+4\dd\phi\wedge\ast_{10}\dd\phi) \nn \\&&
-\frac{1}{8\kappa_{10}^2}\int_{\mathbf{R}^{1,3}\times Y}\;\left(2e^{-2\phi} H\wedge\ast_{10}H+\sum_{p=2,4,6,8}{G^{(p)}\wedge \ast_{10} G^{(p)}}\right).
\label{sugraAct10D}
\end{eqnarray}
Here $\hat{R}$ is the (10D) curvature scalar, $H=\dd B^{(2)}$, $\kappa_{10}^2=\frac{l_s^8}{4\pi}$ is the ten-dimensional gravitational coupling constant with $l_s=2\pi\sqrt{\alpha'}$ the string length, while $G^{(p)}$ is defined in (\ref{dualityG}). In these conventions the coefficient functions $\hat{g}_{MN},\; B^{(2)}_{MN},\; C^{(p)}_{M_1\ldots M_p}$ etc. of all the fields are dimensionless, which means that the field strengths have a dimension of inverse length.\\
We could now insert Lagrangian multipliers to implement the relations of (\ref{dualityG}) and integrate out the redundant fields (see e.g. \cite{bergshoeffIIAsugra, dallAgata, louisMicu, gukovHaack}). This would give us the low-energy effective bulk action as computed in \cite{grimmOrientifold}. However, as additional couplings of the fields will arise from the Chern-Simons action of the D6-branes, we postpone this procedure until we have the full pseudo-action. \\
\indent Now we expand the ten-dimensional curvature scalar using the K\"ahler structure and complex structure moduli \cite{moduliCandelas} and perform a Weyl-rescaling of the metric to the Einstein-frame
\begin{equation}\hat{g}_{MN}\rightarrow e^{\frac{\phi}{2}}\hat{g}_{MN}.\label{rescale}\end{equation}
Under a general rescaling $g\rightarrow\Omega^{-2}g$ of the metric in D dimensions the transformation of the curvature scalar term is \cite{sugraLagBodner, ferraraDimRed}
\begin{equation}
\sqrt{\det g}\, \Omega^{D-2}\, R\rightarrow \sqrt{\det g}\, (R+(D-1)(D-2)(\partial_{\mu}\log{\Omega})^2) .\label{rescaleGen}\end{equation}
Finally, under (\ref{rescale}) the Hodge star of a p-form $\lambda$ transforms as $\ast_{10} \lambda \rightarrow e^{\frac{5-p}{2}\phi}\ast_{10} \lambda$.\\
\indent Next we insert the expansions of the fields and perform the integration over the internal space as in \cite{sugraLagBodner, ferraraDimRed, jockersD7, grimmOrientifold}. The reduction of the curvature scalar is performed in more detail in appendix \ref{reduxR}. Finally we rescale the 4D metric as
\begin{equation}\eta_{\mu\nu}\rightarrow \frac{l_s^6}{V_Y}\eta_{\mu\nu}\label{rescale4d}\end{equation}
where $V_Y$ is the volume of the internal Calabi-Yau manifold $Y$ (in the Einstein frame) and we use equation (\ref{rescaleGen}) as well as the relation between the 4d and 10d gravitational coupling constants $\frac{1}{\kappa_4^2}=\frac{l_s^6}{\kappa_{10}^2}$.
The resulting 4d low energy effective action is

\begin{eqnarray}
S_{IIA}^{(4)}&=&
-\frac{1}{2\kappa_4^2}\int_{\textbf{R}^{1,3}}\left[R\ast_41+\frac{1}{2}\dd\phi\wedge\ast_4\dd\phi+2G_{KL}\dd q^K\wedge\ast_4\dd q^{L} \right.\nn\\ &&
+2\left(G_{ab}+\frac{9}{4}\frac{\mathcal{K}_a\KK_b}{\KK^2}\right)\dd v^a\wedge\ast_4\dd v^b
 +2e^{-\phi}G_{ab}\dd b^a\wedge\ast_4\dd b^b
 \nn \\&&
 +\frac{\KK}{96}e^{-\frac{3}{2}\phi}G^{ab}\left(\dd d_{(3)a}-\KK_{acd}\dd b^c\wedge\tilde{c}^d_{(3)}\right)\wedge\ast_4\left(\dd d_{(3)b}-\KK_{bef}\dd b^e\wedge\tilde{c}^f_{(3)}\right)
 \nn \\&&
 +\left(\frac{\KK}{6}\right)^3e^{-\frac{\phi}{2}}G_{ab}\left(\dd\tilde{c}_3^a -\dd b^a\wedge c_3\right)\wedge\ast_4\left(\dd\tilde{c}_3^b-\dd b^b\wedge c_3\right)\nn \\&&
 +\frac{3}{8\KK}e^{-\frac{\phi}{2}}G^{\alpha\beta}\left(\dd U_{\alpha}-\KK_{a\alpha\beta}\dd b^a\wedge V^{\beta}\right)\wedge\ast_4 \left(\dd U_{\beta}-\KK_{b\beta\gamma}\dd b^b\wedge V^{\gamma}\right)
\nn \\&&
+\frac{\KK}{24}e^{-\frac{\phi}{2}}A^{\hat{K}\hat{L}}\dd \rho_{(2)\hat{K}}\wedge\ast_4 \dd\rho_{(2)\hat{L}}+\frac{3}{2\KK}e^{\frac{\phi}{2}}A_{\hat{K}\hat{L}}\dd \xi^{\hat{K}}\wedge\ast_4\dd \xi^{\hat{L}}
\nn \\&&\left.
+\frac{1}{4}\left(\frac{\KK}{6}\right)^3e^{\frac{\phi}{2}}\dd c_3\wedge\ast_4\dd c_3
 +\frac{\KK}{6}e^{\frac{\phi}{2}}G_{\alpha\beta}\dd V^{\alpha}\wedge\ast_4 \dd V^{\beta} \right] .
\label{bulkAction}
\end{eqnarray}
In the above we have used the (dimensionless) triple intersection numbers of the Calabi-Yau manifold $Y$\footnote{The indices $A, B$ stand for either for $a, b \in\{1,\ldots,h^{(1,1)}_-\}$ or for $\alpha, \beta \in \{1,\ldots,h^{(1,1)}_+\}  $. }
\begin{equation}\KK_{ABC}=\frac{1}{l_s^6}\int_Y \omega_A\wedge\omega_B\wedge\omega_C\end{equation}
as well as their contractions with the Einstein-frame K\"ahler form (which differs by the conformal factor from the string frame version used in (\ref{fieldExp}): $J_E=J_{sf}e^{-\frac{\phi}{2}}$). Note that since the volume form is odd under $\sigma^*$ only integrals over odd 6-forms can be nonvanishing and hence the nonvanishing contractions are
\begin{equation}\begin{array}{ll}
\KK = \frac{1}{l_s^6}\int_Y J_E\wedge J_E\wedge J_E, & 
\KK_a = \frac{1}{l_s^6}\int_Y \omega_a \wedge J_E \wedge J_E, \\\\
\KK_{ab} =\frac{1}{l_s^6}\int_Y \omega_a \wedge\omega_b\wedge J_E, &
\KK_{\alpha\beta}=\frac{1}{l_s^6}\int_Y \omega_{\alpha}\wedge\omega_{\beta}\wedge J_E, \\\\
\KK_{\alpha\beta a}=\frac{1}{l_s^6}\int_Y \omega_{\alpha}\wedge\omega_{\beta}\wedge\omega_a, &
\KK_{abc}=\frac{1}{l_s^6}\int_Y \omega_a\wedge\omega_b\wedge\omega_c.
\end{array}\end{equation}
With these abbreviations we may write
\begin{equation}
\omega_a\wedge\omega_b=\KK_{abc}\tilde{\omega}^{c}, \qquad \omega_a\wedge\omega_{\alpha}=\KK_{a\alpha\beta}\tilde{\omega}^{\beta}, \qquad \omega_{\alpha}\wedge\omega_{\beta}=\KK_{a\alpha\beta}\tilde{\omega}^a. \label{relOmegas}\end{equation}
From now on we will only use the Einstein-frame K\"ahler form and omit the index (note that the fields appearing in (\ref{bulkAction}) are already the Einstein-frame fields as they were extracted from the curvature scalar after the rescaling of the metric, i.e. with the Einstein-frame curvature). $\KK$ is proportional to the volume of $Y$ measured in units of $l_s^6$:,
\bea
\label{volKappa}
\KK=\frac{6}{l_s^6}V_Y.
\eea
\indent The metric on the $\mathcal{N}=1$ moduli space of complex structure deformations is the restriction of the $\mathcal{N}=2$ metric to a Lagrangian submanifold of the $\mathcal{N}=2$ moduli space. Before orientifolding the complex structure moduli span a quaternionic K\"ahler manifold with a K\"ahler potential given by \cite{moduliCandelas, sugraLagBodner, grimmOrientifold}
\begin{equation}\label{K_CS} K_{CS}=-\log{\Big(\frac{i}{l_s^6}\int_Y \Omega\wedge\bar{\Omega}\Big)}.\end{equation}
The $\mathcal{N}=1$ complex structure moduli space is not necessarily a complex manifold so in this sense we cannot really call the metric on it K\"ahler, but it is still given by the Hessian matrix of $K_{CS}$ restricted to $\mathcal{M}_{\mathcal{N}=1}^{CS}$
\begin{equation}
G_{KL}=-\frac{\int_Y \chi_K\wedge\bar{\chi}_L}{\int_Y \Omega\wedge\bar{\Omega}}=\frac{\partial^2}{\partial q^K\partial q^L} K_{CS}(q).\label{CSmetric}\end{equation}
\indent We have also defined the metric
\begin{equation}G_{AB}=\frac{1}{4V_Y}\int_Y \omega_A\wedge\ast_6\omega_B =\frac{3}{2l_s^6\mathcal{K}}\int_Y \omega_A\wedge\ast_6\omega_B=-\frac{3}{2}\Big(\frac{\KK_{AB}}{\KK}-\frac{3}{2}\frac{\KK_A\KK_B}{\KK^2}\Big),\label{Gab}\end{equation}
while $G^{AB}$ denotes the inverse matrix. 
As we will show the moduli space of K\"ahler structure deformations is again a K\"ahler manifold spanned by the fields $t^a=\BB^a + ie^{\frac{\phi}{2}}v^a$ with metric $e^{-\phi}G_{ab}$ and K\"ahler potential \cite{moduliCandelas} 
\begin{equation}
\label{K_KS}
K_{KS} = -\log \Big(\frac{4}{3}e^{\frac{3}{2}\phi}\KK \Big).
\end{equation}
 We have also used the identity
\begin{equation}\frac{1}{l_s^6}\int_Y \tilde{\omega}^A\wedge\ast_6\tilde{\omega}^B=\frac{3}{2\KK}G^{AB},\end{equation}
which follows immediately from (\ref{baseNormalization}).\\
\indent Finally, the kinetic terms of the scalars $\xi$ and their dual 2-forms have the prefactors
\begin{equation}
A_{\hat{K}\hat{L}}=\frac{1}{l_s^6}\int_Y \alpha_{\hat{K}}\wedge\ast_6 \alpha_{\hat{L}},\quad \qquad A^{\hat{K}\hat{L}}\equiv(A^{-1})^{\hat{K}\hat{L}}=\frac{1}{l_s^6}\int_Y \beta^{\hat{K}}\wedge\ast_6 \beta^{\hat{L}}.
\label{defA_KL}\end{equation}
They fulfill $\ast_6 \alpha_{\hat{K}}=A_{\hat{K}\hat{L}}\beta^{\hat{L}}$.

\section{The D6-brane Action}
\label{brane}
We now add a D6-brane to the theory in such a way that the $\mathcal{N}=1$ supersymmetry of the bulk theory remains unbroken. The restrictions imposed by this requirement are discussed in section \ref{susyCond}. In section \ref{braneModuli} we discuss the massless moduli space of the brane and its relation to the moduli controlling the complex and K\"ahler structures of the ambient space. Using the formulae for pullbacks to the brane world-volume derived in appendix \ref{pullbacks} we then proceed to integrate out the internal cycle wrapped by the brane in the Dirac-Born-Infeld and Chern-Simons actions. Finally, in section \ref{tadpoles} the tadpole cancellation conditions on the cycles wrapped by the brane and orientifold plane are briefly reviewed for completeness.

\subsection{Supersymmetric cycles and calibration conditions}
\label{susyCond}
\noindent In addition to the action describing the low-energy dynamics of the closed string sector derived in the previous section we would like to include open string fields by allowing for a D6 brane in such a way that the $\mathcal{N}=1$ supersymmetry of the bulk orientifold theory is preserved. The easiest way to describe this brane on the orientifold background is to consider first a brane wrapped on a cycle $\Pi_0$ of the Calabi-Yau manifold $Y$ and add a mirror brane, wrapped on the mirror cycle $\Pi_0'=\sigma_* \Pi_0$. For the Poincar$\acute{\textnormal{e}}$-dual 3-forms $\pi_0,\; \pi_0'$ of these cycles the fact that the volume form of $Y$ is odd under $\sigma^*$ then implies that $\pi_0'=-\sigma^*\pi_0$. We can expand these forms into the basis of 3-forms defined in \ref{basisforms} as
\begin{equation}\pi_0=\pi_0^{\hat{K}}\alpha_{\hat{K}} + \pi_{0\;\hat{L}}\beta^{\hat{L}};\qquad \pi_0'=-\pi_0^{\hat{K}}\alpha_{\hat{K}} + \pi_{0\;\hat{L}}\beta^{\hat{L}}.
\end{equation}
The forms describing the branes and mirror branes together are $\pi_{\pm}=\frac12 (\pi_0\pm\pi_0')$\footnote{The inclusion of the factor $\frac12$ into this definition means that we will not have to divide by 2 when computing the DBI and CS actions for the brane/image brane pair.} with
\begin{equation}
\pi_-=\pi_-^{\hat{K}}\alpha_{\hat{K}}=\pi^{0\hat{K}}\alpha_{\hat{K}},\qquad \pi_+=\pi_{+\hat{K}}\beta^{\hat{K}}=\pi_{0\hat{K}}\beta^{\hat{K}}.\end{equation}

\indent For simplicity we assume that there are no fixed points under $\sigma$ on the cycle $\Pi_0$, as this would result in the appearance of additional twisted massless states \cite{branesAtAngles}. In this case the low energy dynamics of the open strings ending on a stack of $N$ branes can be described by a $U(N)$ Yang-Mills gauge theory on the world volume, involving a one-form gauge field $\hat{A}$ as well as a set of scalar fields in the adjoint representation. As we are only considering the abelian case of a single brane the scalars are not charged with respect to the gauge field. These scalars have a direct geometrical interpretation in that they are the moduli describing possible fluctuations of the brane in the directions normal to its world-volume. The Kaluza-Klein reduction of the massless modes of the gauge boson can be carried out in a manner similar to the bulk fields, i.e. by expanding it into harmonics on $\Pi_+$ and then projecting out some of the resulting degrees of freedom by the orientifold action. However, finding a suitable description for the second type of moduli describing the fluctuations of the brane turns out to be much less straightforward. This is because the requirement that the brane should preserve $\mathcal{N}=1$ supersymmetry imposes several restrictions on the possible allowed fluctuations of the brane, which we will now discuss. 

\indent As is well-known these constraints arise from the fact that the new fermionic fields introduced by the open strings ending on the brane are not necessarily invariant under an infinitesimal supersymmetry variation induced by one of the two SUSY generators that are left unbroken by compactification on a Calabi-Yau manifold. However, under certain conditions the additional kappa symmetry of the worldsheet action may be used to cancel the fermion variation induced by a certain linear combination of the original SUSY generators. This linear combination is then the generator of the unbroken $\mathcal{N}=1$ supersymmetry in 4 dimensions.\\
\indent It can be shown that for a pair of D6 branes wrapped on a cycle $\Pi_+$ the supersymmetry condition is that $\Pi_+$ is a special Lagrangian (sLag) manifold calibrated by $e^{i\theta}\iota^*\Omega$ together with a certain condition on the gauge flux \cite{5branesMembranes_BeckerStrom, susypbranes_Strominger, susy3cycles_Kachru}. After rescaling to the Einstein frame the calibration conditions read
\begin{equation}
\iota^*(e^{\frac{\phi}{2}}J)=0,\qquad e^{\frac{1}{2}(K_{CS}-K_{KS})}\iota^*\Omega=e^{i\theta}e^{\frac{3}{4}\phi}\mathrm{dvol}|_{\Pi_+}^E,\qquad  \iota^*B^{(2)}-\frac{l_s^2}{2\pi} \hat{F}|_{\Pi_+} =0.\label{calibration}
\end{equation}
$\hat{F}$ is the field strength of the $U(1)$ gauge field on the brane world-volume and the notation $\hat{F}|_{\Pi_+} $ denotes the internal part of $\hat F$, while $K_{CS}$ and $K_{KS}$ were defined in (\ref{K_CS}) and (\ref{K_KS}), respectively. Note that the second calibration condition is scale invariant as a rescaling of $\Omega\rightarrow e^{-\mathrm{Re}\;h}\Omega$ is exactly compensated by the resulting shift of the K\"ahler potential $K_{CS}\rightarrow K_{CS}+2\mathrm{Re}\;h$. The phase $\theta$ parametrizes which of the original supersymmetry generators is left unbroken by adding this brane and flux configuration. In the orientifold theory that we are considering this phase is not arbitrary, as the operation of orientifolding itself breaks the supersymmetry from $\mathcal{N}=2$ to $\mathcal{N}=1$. To ensure that both operations leave the same supersymmetry unbroken the sLag cycles wrapped by the branes must be calibrated with respect to the same form as the sLag cycle of the O6 orientifold plane at the fixed point locus of the orientifold involution $\sigma$, i.e. the $\theta$'s of equations (\ref{calibration}) and (\ref{pullbackOmega}) must be equal.

 \indent The D6-brane may carry gauge flux $f$ arising as the vacuum expectation value of the field strength,
\begin{equation}\hat{F}=f+\dd \hat{A}. \end{equation}
To preserve Lorentz invariance of the 4D theory these background fluxes must be forms on the internal part of the world-volume, i.e. the cycle $\Pi_+$.
Note that as we are restricting ourselves to the massless modes of the gauge field $\dd\hat{A}$ contains no internal two-forms, so we actually have $\hat{F}|_{\Pi_+}=f$.
By the Freed-Witten anomaly cancellation condition \cite{Freed:1999vc} the gauge flux $f$ must be quantised according to 
\bea
\frac{l_s^2}{2\pi} f  + \frac{1}{2}\, w_2(\Pi_+) \in H^2(\Pi_+,\mathbb Z).
\eea 
Here $w_2(\Pi_+)$ denotes the second Stieffel-Whitney class of $\Pi_+$, which vanishes for sLags on a Calabi-Yau 3-fold due to triviality of the normal bundle \cite{Bryant:1998wv}. Hence $\frac{l_s^2}{2\pi} f$ must be integer quantised.
Supersymmetric fluxes are furthermore subject to the third condition in eq. (\ref{calibration}), by which the pullback of the $B^{(2)}$-field must cancel the gauge flux.
Integrality of $f$ thus precludes the possibility of switching on discrete $B_+^{(2)}$-field along the D6-brane, 
\bea
 \iota^* B^{(2)}_+ =0.
 \eea 
Furthermore the supersymmetric flux must be writable as the pullback of forms defined globally on $Y$: $f=f^a\iota^*\omega_a$.
This allows us to combine the flux components with the bulk $B^{(2)}$ field into \begin{equation}\BB=\BB^a(x)\, \omega_a=(b^a(x)- \frac{l_s^2}{2\pi} f^a)\, \omega_a.\end{equation}
In the supersymmetric vacuum the  calibration condition then enforces $\iota^*\BB=0$.

\subsection{The open string moduli}
\label{braneModuli}
\noindent Next we turn to the restrictions these calibration conditions impose on the $\mathcal{N}=1$ moduli space $\MM$. The position in the ambient Calabi-Yau manifold of the fluctuating brane can be described at least for small fluctuations by normal vector fields on the world-volume 
\bea
\mathcal{W} = \mathbf{R}^{1,3} \times \Pi_+
\eea
via the exponential map. Let $\Phi:\mathcal{W}\rightarrow T\Pi_+$ be a normal vector field such that the position of the displaced brane is given by $\mathcal{W}_{\Phi}=\left\{\exp_{\Phi}(p)\equiv\exp_p(\Phi(p));\; p\in\mathcal{W}\right\}$. 
This exponential map is described in more detail in appendix \ref{pullbacks}.
To obtain a supersymmetric theory we must allow only such fluctuations that respect the calibration conditions (\ref{calibration}) with $\iota$ replaced by $\exp_{\Phi}$. Clearly, this space of allowed deformations then depends on the bulk moduli through the calibration conditions. In other words, the true moduli space $\MM$ does not have a simple product structure,
\begin{equation}\MM\neq \mathcal{M}_{bulk}\times \mathcal{M}_{brane}.\label{modSpace}\end{equation}
\indent The combined bulk and brane moduli space can be described mathematically with the help of the concept of relative cohomology and the variation of relative Hodge structures \cite{WittenTFT, ChiralRingsCohomology, LercheMayr_SpecGeomOpenClosed}. However, this abstract formulation makes it difficult to explicitly perform the dimensional reduction of the action. In the limit of small fluctuations $\delta v^a$, $\delta \BB^a$ etc. around a background value and small deformations of the brane we can \emph{approximate} the moduli space by the product of eq. (\ref{modSpace}). Here we will follow the approach taken in refs. \cite{grimmD5, jockersD7} in the IIB case by computing this approximation to the effective action and then attempting to find a corresponding K\"ahler potential.
One could then try to use more advanced techniques to extrapolate this result to a K\"ahler potential
 on the true moduli space which reduces to our result in the limit of small fluctuations. However, this is beyond the scope of this note and will not be pursued here. \\
\indent Let us therefore choose a definite reference point on the relevant part of the bulk moduli space given by $J_0$, $\BB_0$ and $\Omega_0$ and impose the calibration condition (\ref{calibration}) in terms of these fields. In particular this means that we require the brane displacements to respect the sLag conditions with respect to these background values of the K\"ahler and holomorphic 3-forms, i.e. we are looking for the moduli space of sLag manifolds in a fixed background Calabi-Yau manifold.\\ 
\indent This moduli space has been shown by McLean to be a smooth (real) manifold, the dimension of which is given by the first Betti number of $\Pi_+$ \cite{McLean_Slag}. Let us recall the arguments: As we stated before, small deformations of the cycle may be described by normal vector fields using local tubular coordinates. By the vanishing of the K\"ahler form on $\Pi_+$ (eq. \ref{calibration}), the complex structure restricted to $N\Pi_+\subset TY$ maps the normal space isomorphically to $T\Pi_+$. Using the induced metric we therefore also obtain an isomorphism to the cotangent space of $\Pi_+$. Concretely this isomorphism is simply given by the interior product of the normal vector field with the K\"ahler form. Finally, the conditions of eq. (\ref{calibration}) applied to the sLag manifold imply that the corresponding one-form on $\Pi_+$ is both closed and co-closed and hence harmonic. This yields McLean's result quoted above (see also e.g. \cite{GrossJoyce_CYGeom, Hitchin_Slag, Salur_SlagDef, Salur_Fred} for further details). \\
\indent The discussion above applies to deformations of sLag manifolds embedded in general Calabi-Yau manifolds. We must now also take into account the effect of the orientifold projection on the spectrum. As the worldsheet parity operation on the open strings flips the sign of the derivatives tangent to the brane while leaving those normal to the brane invariant \cite{polchinskiOrientifold}, we deduce that the normal vector fields describing the fluctuations of the brane in an orientifold must have even parity under pushforward along $\sigma$. This also makes sense geometrically, as viewed in the original CY manifold brane and image brane should fluctuate in unison. Due to the negative parity of the K\"ahler form this implies that the one-forms associated with the allowed fluctuations have negative parity under $\sigma^*$, as they are constructed by contracting the K\"ahler form with the corresponding normal vector.\\
\indent Let us choose a basis $A^I,\; I=1,\ldots, b_1^-(\Pi_+)$, of the space of negative parity harmonic one-forms on $\Pi_+$ and the corresponding sections of the normal bundle $s^I$ satisfying 
\begin{equation}A^I=\iota^*(e^{\frac{\phi}{2}}i_{s^I}J_0), \qquad I=1,\ldots, b_1^-(\Pi_+). \label{relAS}\end{equation}
As we have just shown we can expand the normal vector describing the sLag deformations of the brane into this basis as 
\begin{equation}\Phi= \Phi_I(x, \zeta) \, s^I\; \textnormal{resp. in coordinates }\Phi^m= \Phi_I(x, \zeta) \,  (s^I)^m.\end{equation}
Here $x$ and $\zeta$ denote coordinates along $\mathbf R^{1,3}$ and $\Pi_+$ respectively, and the index $m$ labels the coordinates $y^m$ normal to the brane.

In the course of the Kaluza-Klein reduction of the action we then drop the dependence of the scalar fields $\Phi$ on the internal coordinates, so the deformation moduli of the brane contribute the set $\Phi_I(x),\; I=1,\ldots, b_1^-(\Pi_+)$ of real scalar fields to the effective theory.\\
\indent In exactly the same manner as for the K\"ahler form (explained e.g. in \cite{Hitchin_Slag}) it is possible to use the other two calibration conditions to find further relations between the pullbacks of the bulk basis forms and the $A^I$. For example, the condition $\exp_{c_I s^I}^* \BB_0=0$ for constants $c_I$ shows that the supersymmetric fluctuations must obey $\dd \iota^*(i_{s^I}\BB_0)=0$. This can actually also be seen directly using the pullback formula derived in appendix \ref{pullbacks}, which to second order gives
\begin{equation}0=\exp_{c_Is^I}^* \BB_0= \iota^*\BB_0 + c_I\dd\iota^*(i_{s^I}\BB_0)+c_Ic_J\dd\iota^*(i_{s^I}\dd i_{s^J}\BB_0)+ \ldots\end{equation}
The unique harmonic representative of the cohomology class of $\iota^*(i_{s^I}\BB_0)$ may then be expanded into the basis of harmonic forms introduced above, such that
\begin{equation}\iota^*(i_{s^I}\BB_0)=\dd r^I + M^I_J A^J \label{exps_IB}\end{equation}
where the $r^I$ are functions on $\Pi_+$ and the matrix $M^I_J$ is real as both $\BB$ and $A^I$ are real.\\
\indent The second set of fields introduced by the D6-brane come from the reduction of the abelian gauge boson $\hat{A}$. This effective field enters the open-string interactions in which the vertex operators contain derivatives tangent to the brane. As explained above this requires the surviving fields in the orientifolded theory to have negative parity under pullback along $\sigma$. This implies that we may expand it using the above basis as 
\begin{eqnarray}
\hat{A}(x,y)&=& \frac{\pi}{l_s}A(x)P_-(y)+\frac{\pi}{l_s}a_I(x)A^I(y), \nn \\
P_-(y)&=&\genfrac{\{.}{}{0pt}{}{\; 1, \qquad  y \in \Pi_0} {-1, \qquad y \in \Pi_0'}.
\label{expA}
\end{eqnarray}
$A(x)$ describes a 4-dimensional $U(1)$ gauge boson, while the $a_I$ are Wilson line moduli. We have extracted a factor of $\frac{1}{l_s}$ so that $A_{\mu}$, $a_I$ and $A^I_a$ are all dimensionless, while the factor of $\pi$ has been chosen for later convenience. \\
\indent The fact that the deformation moduli and the Wilson line moduli are in 1-1 correspondence is no accident \cite{Lerche_SpecGeom, MayrLerche_MirrorSymm}. Rather, we know that the resulting 4D theory must be an $\mathcal{N}=1$ supergravity, which in particular implies that the real moduli fields must pair up into complex fields representing the bosonic part of a chiral supermultiplet. We have already seen this in the case of the K\"ahler structure moduli which pair up with the fields from the reduction of the Kalb-Ramond field and parametrize the complexified K\"ahler cone. The fact that this must also work for the brane moduli and that the first Betti number of $\Pi_+$ may be odd forces the deformation and Wilson line moduli to match up in the manner described above.\\
\indent Note that this situation is quite different from the case of space-filling D-branes in a IIB theory. In that case the cycles wrapped by the branes are holomorphic submanifolds, and the Wilson line and deformation moduli are complex fields individually. In fact, for D7- and D5-branes they turn out to be correct chiral coordinates of the resulting supergravity theory, at least in the limit where they may be described independently of the complex structure deformations \cite{grimmD5, jockersD7}.

\subsection{The abelian world-volume action}
\noindent The low energy dynamics of the D6-brane in the abelian case are described in the string frame by the well known Dirac-Born-Infeld and Chern-Simons actions
\bea
S_{DBI} &=& -\mu_6 \int_{\mathcal{W}} \dd^7x \left[e^{-\phi}\sqrt{-\det\left(\mathrm{P}_{\Phi}\left(\hat{g}^{sf}+B^{(2)}\right)-\frac{l_s^2}{2\pi}\hat{F}\right)}\right],
\label{DBI}\eea
\bea
S_{CS} &=& \mu_6 \int_{\mathcal{W}} \left(\mathrm{P}_{\Phi}\left(\sum_n C^{(n)} e^{-B^{(2)}}\right)e^{\frac{l_s^2}{2\pi}\hat{F}}\right), \quad\quad \mu_6 = \frac{2 \pi}{l_s^7}.
\label{CS}\eea
Here $\mathrm{P}_{\Phi}$ is the pullback onto the fluctuating world-volume. We introduce local tubular coordinates around the rest-position $\iota:\mathcal{W}\hookrightarrow Y$ of the branes with $a, b$ labelling directions along and $n, m$ perpendicular to the branes. The position of the fluctuating brane in the transverse directions is then given by $y^m=\frac{l_s}{2}\Phi^m$. Note that the $\Phi^m$ are dimensionless as we have extracted a factor of $l_s$. The terms of the pullbacks relevant for the reduction of these actions are computed in appendix \ref{pullbacks}. These actions form the effective Lagrangian describing open string interactions at leading order in derivatives and at tree level in the string coupling.\\
\indent Note that as always the Kalb-Ramond two-form field $B^{(2)}$ and the world-volume gauge field strength appear together 
such that a gauge transformation of the $B^{(2)}$-field may be canceled by a corresponding transformation of the gauge field \cite{TASIpolchinski}. 
The relative sign in $B^{(2)}-\frac{l_s^2}{2\pi}\hat{F}$ is the one consistent with the conventional sign choice of eq. (\ref{dualityG}); it turns out that this sign will allow us to perform the dualization procedure consistently.\footnote{We could just as well have used the convention $B^{(2)}+\frac{l_s^2}{2\pi}\hat{F}$, however then we would have also had to define the field strengths as $G^{(p)}:=\mathrm{d}C^{(p-1)}+\mathrm{d}B^{(2)}\wedge C^{(p-3)}$}\\

\subsubsection{Reduction of the DBI action}
\label{DBIsection}
\noindent We are now ready to determine the 4-dimensional effective action resulting from the DBI action up to second order in derivatives by plugging the expressions for the pullbacks computed in appendix \ref{pullbacks} into the action of equation (\ref{DBI}). In general, contractions, Hodge stars or volume forms are evaluated using the Einstein frame metric. If we use the string frame metric this will be indicated explicitly.\\
To obtain the action in terms of the moduli we expand the determinant up to second order in derivatives (or equivalently second order in the string length $l_s$) and fluctuations around the background values of the various moduli fields. To this end we define the matrix $\mathcal{X}$ by $\iota^*\hat{g}^{sf}+\mathcal{X}=\mathrm{P}_{\Phi}\left(\hat{g}^{sf}+B^{(2)}\right)-\frac{l_s^2}{2\pi}\hat{F}$ 
and use the Taylor expansion of the determinant
\begin{eqnarray}
\sqrt{-\det (\iota^*\hat{g}^{sf}+\mathcal{X})}&=&\sqrt{-\det \iota^*\hat{g}^{sf}}\left[1+\frac{1}{2}\Tr((\iota^*\hat{g}^{sf})^{-1}\mathcal{X})\right.\nn \\&&\left.
+\frac{1}{8}\left[(\Tr((\iota^*\hat{g}^{sf})^{-1}\mathcal{X}))^2-2\Tr((\iota^*\hat{g}^{sf})^{-1}\mathcal{X})^2\right]+\ldots\right].
\label{taylorDet}\end{eqnarray}

The result of inserting the explicit expressions for the pullbacks of the metric and
the Kalb-Ramond field into the expansion of the determinant is given in the string frame by
\begin{eqnarray}
\label{DBIa}
S_{DBI}&=&-\mu_6\int_{\mathcal{W}}\dd^7 xe^{-\phi}\sqrt{-\det \iota^*\hat{g}^{sf}}\left[1+\frac{1}{8}l_s^2 \partial_{\mu}\Phi_I\partial^{\mu}\Phi_J \hat{g}(s^I,s^J)\right.\nn \\&&
+\frac{1}{16}l_s^2 F_{\mu\nu}F^{\mu\nu}  +\frac{1}{8}l_s^2 \Phi_I\Phi_J\hat{g}^{ab}\hat{g}_{mn}(s^I)^m_{,a}(s^J)^n_{,b}
\nn \\&&
+\frac{1}{8}l_s^2\partial_{\mu}\Phi_I\partial^{\mu}\Phi_J \hat{g}^{ab}(i_{s^I}\BB_0)_a(i_{s^J}\BB_0)_b +\frac{1}{4}\BB_{ab}\BB^{ab}\nn \\&&
\left. -\frac{1}{4}l_s^2\partial_{\mu}\Phi_I\partial^{\mu}a_J \hat{g}^{ab}(i_{s^I}\BB_0)_a A^J_b+\frac{1}{8}l_s^2\partial_{\mu}a_I\partial^{\mu}a_J \hat{g}^{ab}A^I_a A^J_b
+\ldots\right].
\end{eqnarray}
In the above we have dropped any terms with 3 or more factors of derivatives of the brane fields or fluctuations. As we have already explained, the action computed by treating the bulk and brane moduli separately only holds in the limiting case of small fluctuations and the expansion of the DBI action is valid for small derivatives of the brane fields. This has allowed us to replace $\BB\rightarrow\BB_0$ in the terms already including two derivatives. \\
The block diagonal form of the metric $\hat{g}$ allows us to write (cf. (\ref{rescale4d}) for Weyl rescaling)
\begin{equation}\dd^7x \sqrt{-\det \iota^*\hat{g}^{sf}}=\Big(\frac{6}{\KK}\Big)^2e^{\frac{7}{4}\phi}\mathrm{dvol}|_4^E\wedge\mathrm{dvol}|_{\Pi_+}^E. \end{equation}
Our conventions will be that for p-forms $\eta$, $\lambda$ on either $\textbf{R}^{1,3}$ or $\Pi_+$ there holds
\begin{equation}\frac{1}{p!}\eta_{a_1\ldots a_p}\lambda^{a_1\ldots a_p}\mathrm{dvol}=\eta\wedge\ast\lambda.\end{equation}
Finally, one can use the relation (\ref{relAS}) and the fact that the complex structure squares to -1 to show that $\hat{g}^{sf}_{nm}(s^I)^n(s^J)^m=e^{-\phi}(\hat{g}^E)^{ab}A^I_aA^J_b$ and \begin{equation}(\hat{g}^{sf})^{ab}(\hat{g}^{sf})^{cd}(\dd A^I)_{ac}(\dd A^J)_{bd}=\frac{1}{2}(\hat{g}^{sf})^{ab}\hat{g}^{sf}_{mn}(s^I)^m_{,a}(s^J)^n_{,b}.\end{equation}

\indent Using the identities above and evaluating the integrals over the internal cycle we obtain up to second order after rescaling to the Einstein frame
\begin{eqnarray}
S^E_{DBI}&=&-\mu_6l_s^3\int_{\textbf{R}^{1,3}}\left[ \frac{3}{4\KK}l_s^2 e^{-\frac{\phi}{4}}\mathcal{H}^{IJ}\left(\dd \Phi_I\wedge\ast_4\dd \Phi_J +\dd a_I\wedge\ast_4 \dd a_J\right.\right.\nn \\& & \left.
+M^L_IM^K_J\dd\Phi_L\wedge\ast_4\dd\Phi_K-2 M^L_I\dd\Phi_L\wedge\ast_4\dd a_J\right) \nn \\&&
\left.+\frac{1}{8}l_s^2e^{-\frac{\phi}{4}}V_{\Pi_+}F\wedge\ast_4 F  + \frac{18}{\KK^2}e^{-\frac{\phi}{4}}\mathcal{N}_{ab}\BB^a\BB^b\ast_4 1 + \frac{36}{\KK^2}e^{\frac{3}{4}\phi}V_{\Pi_+}\ast_4 1
 \right]
\label{DBIaction}\end{eqnarray}
with the abbreviations
\begin{eqnarray}
V_{\Pi_+}&=&\frac{1}{l_s^3}\int_{\Pi_+}\mathrm{dvol}|^E_{\Pi_+} \stackrel{(\ref{calibration})}{=} \frac{1}{l_s^3}e^{\frac{1}{2}(K_{CS}-K_{KS})}e^{-\frac{3}{4}\phi}e^{-i\theta}\pi_{0\hat{K}}\int_Y \Omega\wedge \beta^{\hat{K}}, \nn \\
\mathcal{H}^{IJ}&=&\frac{1}{l_s^3}\int_{\Pi_+} A^I\wedge \ast_{\Pi_+} A^J, \nn\\
\mathcal{N}_{ab}&=&\frac{1}{l_s^3\pi^2}\int_{\Pi_+} i^*\omega_a\wedge\ast_{\Pi_+} i^*\omega_b
\label{coeffDBI}\end{eqnarray}
and  $M^I_J$ defined in eq. (\ref{exps_IB}).  
The Dirac-Born-Infeld action provides the kinetic terms for the world-volume gauge field as well as the Wilson line and transverse displacement moduli. Finally, the last two terms represent NS-NS tadpoles, which will be cancelled by the corresponding terms of the orientifold action as we will see in section \ref{tadpoles}.

\subsubsection{Reduction of the Chern-Simons action}
\label{CSreduction}
\noindent The reduction of the Chern-Simons action from equation (\ref{CS}) proceeds in a similar manner. The forms whose pullback to the world-volume appear in the action (\ref{CS}) can all be written as $\omega\wedge\eta$, where $\omega$ is a differential form on 4d spacetime and $\eta$ is a closed form on $Y$. For such forms, the pullback may be written as
\begin{eqnarray}
P_{\Phi} (\omega\wedge\eta)&=&\omega\wedge\iota^*\eta + \frac12 l_s \omega\wedge\dd \Phi_I \wedge\iota^*i_{s^I}\eta +\frac{1}{4}l_s^2\omega\wedge(\Phi_I\dd\Phi_J)\wedge\iota^*(i_{s^J}\dd i_{s^I}\eta) \nn \\&& +\frac{1}{8}l_s^2\omega\wedge\dd\Phi_I\wedge\dd\Phi_J\wedge\iota^*(i_{s^J}i_{s^I}\eta)+\ldots
\label{pullbackForms}\end{eqnarray}
up to terms which vanish upon integration over the internal part of the world-volume. The derivation of this result is presented in appendix \ref{pullbacks}.\\
\indent Using the considerations presented above and keeping only terms up to second order in $l_s$ we obtain the low-energy approximation to the Chern-Simons action encoding the couplings of the massless 4-dimensional bulk and brane fields,
\begin{eqnarray}
S_{CS}&=&\mu_6l_s^3\int_{\textbf{R}^{1,3}} \left[\frac12 l_s \Phi_I\left(\mathcal{M}^{Ia}+\frac{1}{4\pi}\mathcal{T}^{aIJ}\Phi_J\right)\left(\dd d_{(3)a}-\KK_{abc}\dd\BB^b\wedge\tilde{c}^c_{(3)}\right)\right.\nn \\&& -\frac12 l_s \left(\mathcal{M}^{Ia}\Phi_I\KK_{abc}\BB^b-\mathcal{L}_a^I a_I+\frac{1}{4\pi}\mathcal{T}^{aIJ}\KK_{abc}\Phi_I\Phi_J\BB^b\right)\nn\\&&
\times\left(\dd\tilde{c}^c_{(3)}-\dd\BB^c\wedge c_{(3)}\right)
+\frac12 l_s\left(\frac{1}{2}\mathcal{M}^{Ia}\KK_{abc}\Phi_I\BB^b\BB^c-\mathcal{L}_a^I a_I\BB^a\right.\nn \\&&
\left.+\frac{1}{8\pi}\mathcal{T}^{aIJ}\KK_{abc}\BB^b\BB^c\Phi_I\Phi_J\right)\dd c_{(3)}+\frac{1}{8} l_s^5\pi_{0\hat{K}}\xi^{\hat{K}} F\wedge F \nn \\&&
+\frac{1}{4} l_s\left( l_s\mathcal{Q}^{IJ\hat{K}}\dd \Phi_I\wedge \dd a_J
-  \frac{1}{2}l_s \mathcal{P}_a^{IJ\hat{K}}\BB^a\dd\Phi_I\wedge\dd\Phi_J\right.\nn\\&& 
\left. -\pi_0^{\hat{K}} F\right)\wedge \rho_{(2)\hat{K}}+\frac{1}{4}l_s^2 \tilde{\mathcal{M}}^{I\alpha}\Phi_I \dd U_{\alpha}\wedge F \nn \\&& \left. +\frac{1}{4}l_s^2  V^{\alpha}\wedge \left(\tilde{\mathcal{L}}_{\alpha}^I \dd a_I-\KK_{\alpha\beta a} \tilde{\mathcal{M}}^{I\beta}\BB^a \dd \Phi_I\right)\wedge F
\right].
\label{actionCS}\end{eqnarray}
The abbreviations used in the above are
\begin{equation}\begin{array}{rclrcl}
\mathcal{M}^{Ia}&=&\frac{1}{l_s^3}\int_{\Pi_+} \iota^*(i_{s^I}\tilde{\omega}^a),&
\tilde{\mathcal{M}}^{I\alpha}&=&\frac{1}{l_s^3}\int_{\Pi_-} \iota^*(i_{s^I}\tilde{\omega}^{\alpha}), \\\\
\mathcal{L}_a^I&=&\frac{1}{l_s^3}\int_{\Pi_+} (\iota^*\omega_a)\wedge A^I, &
\tilde{\mathcal{L}}^I_{\alpha}&=&\frac{1}{l_s^3}\int_{\Pi_-} (\iota^*\omega_{\alpha})\wedge A^I, \\\\
\mathcal{T}^{aIJ}&=&\frac{1}{l_s^2}\int_{\Pi_+} \iota^*(i_{s^I}\dd i_{s^J} \tilde{\omega}^a),&
\tilde{\mathcal{T}}^{\alpha IJ}&=&\frac{1}{l_s^2}\int_{\Pi_-} \iota^*(i_{s^I}\dd i_{s^J} \tilde{\omega}^{\alpha}),\\\\
\mathcal{Q}^{IJ\hat{K}}&=&\frac{1}{l_s^3}\int_{\Pi_+} (\iota^*i_{s^I}\beta^{\hat{K}})\wedge A^J,&
\mathcal{P}_a^{IJ\hat{K}}&=&\frac{1}{l_s^3}\int_{\Pi_+} \iota^*(i_{s^J}i_{s^I}(\beta^{\hat{K}}\wedge\omega_a)).
\end{array}\label{coeffCS}\end{equation}
Note that in the term $\mathcal{P}_a^{IJ\hat{K}}\BB^a\dd\Phi_I\wedge\dd\Phi_J$ we may replace $\BB^a\rightarrow \BB_0^a$ as it is already of second order in derivatives. Then equations (\ref{exps_IB}) may be used to show that $\BB_0^a\mathcal{P}_a^{IJ\hat{K}}=\mathcal{Q}^{IL\hat{K}}M^J_L-\mathcal{Q}^{JL\hat{K}}M^I_L$. Finally let us note for later use that $\pi_{0\hat{K}}\mathcal{Q}^{IJ\hat{K}}=0$. \\
\indent The Chern-Simons action includes terms potentially inducing kinetic mixing between the gauge fields of the brane and the bulk RR forms as well as a St\"uckelberg mass term for the brane gauge field. However, we must still eliminate the additional degrees of freedom introduced in the form of $C^{(5)}$ and $C^{(7)}$, so we postpone the discussion of the significance of these terms until this has been accomplished in section \ref{dualization}. 

A word of caution is in order concerning the objects $\mathcal{L}_a^I$ (and their counterparts $\tilde{\mathcal{L}}_{\alpha}^I$). Note that the 2-form $\iota^*\omega_a$ is a pullback from the ambient Calabi-Yau onto the brane, while the 1-form $A^I$ does clearly not arise by pullback from the ambient space.
In general a wedge product between two such objects as in $\mathcal{L}_a^I$ can be non-zero. This is not in contradiction with known vanishing results
which are valid in special geometric situations.
For example, if we consider a holomorphic divisor in a Calabi-Yau 3-fold, then the wedge-product between a 2-form pulled back from the ambient space and a 2-form trivial on the ambient space is known to vanish, as discussed in detail in  \cite{jockersD7,Jockers:2005zy}.

\subsubsection{Tadpole cancellation}
\label{tadpoles}
\noindent It is well known that the appearance of R-R tadpoles in a string compactification leads to inconsistencies in the theory. 
From the Chern-Simons action one reads off that our D6-branes may act as sources of such tadpoles via the term $\int_{\mathcal{W}}C^{(7)}$. One of the most important features of orientifold compactifications is the appearance of orientifold planes with negative charge and tension, which may cancel the tadpoles of the D-branes. The coupling of the orientifold plane to the closed string fields is described by a sum of a DBI-type and Chern-Simons-type action which are formally identical to those of the D6-branes after setting the world-volume fields to zero. The orientifold plane is located at the fixed point locus of the antiholomorphic involution $\sigma$. In our case the supersymmetry conditions (\ref{pullbackJ}), (\ref{pullbackOmega}) imply that the theory involves an O6-plane wrapped around a special Lagrangian 3-cycle $\sigma_{O6}$ of $Y$, i.e. with world-volume given by $\Sigma_{O6}=\mathrm{R}^{(1,3)}\times\sigma_{O6}.$ It has been shown that the charge and tension of $O^{(6)}$-planes is -4 times that of a D6-brane \cite{polchinskiRRCharges}. Hence the RR tadpole cancellation condition reads
\begin{equation}\int_{\mathcal{W}}C^{(7)} -2 \int_{\Sigma_{O6}}C^{(7)}=0\label{tadpoleCond}.\end{equation}
This yields the restriction that the sum of the homology classes of the brane and orientifold plane (weighted with the appropriate factor) must cancel \cite{blumenhagen4Dcomp, DbraneModelMarchesano}
\begin{equation}\left[\Pi\right]+\left[\Pi'\right]-4\left[\sigma_{O6}\right]=0.\end{equation}
In the following we will assume that $Y$ and the involution $\sigma$ have been chosen appropriately such that this condition is satisfied.\\
\indent In addition to the R-R tadpoles, also tadpoles for the NS-NS fields may occur. In contrast to the case just discussed such NS-NS tadpoles do not in general render the theory inconsistent \cite{DilatonTadpoles, tadpolesVacuumredef}. They signify that the background one has chosen to expand the theory around is not a proper solution of the ten-dimensional equations of motion. In the effective action such tadpoles appear as scalar potentials for these scalar fields potentially inducing an instability of the vacuum. In section \ref{DBIsection} we saw that the DBI action of the D6-brane introduces tadpoles for the dilaton and the Kalb-Ramond field $\BB$. The Dirac-Born-Infeld part of the orientifold action is given in the string frame by
\begin{equation}S_{O6}=4\mu_6\int_{\Sigma_{O6}}\dd^7 x \, \, e^{-\phi}\sqrt{-\det \iota^*(\hat{g}+B^{(2)})}.\end{equation}
By the Ramond-Ramond tadpole cancellation condition this exactly cancels the last two terms of the brane DBI action of equation (\ref{DBIaction}), so we will drop these terms in the following.

\section{The $\mathcal{N}=1$ characteristic data}
\label{4daction}
After adding up the contributions of bulk and brane to the 4d action we must eliminate the additional degrees of freedom introduced by using the democratic formulation of supergravity. This procedure is carried out in section \ref{dualization}. Having obtained the action in terms of only propagating degrees of freedom we then recast it in the standard form of $\mathcal{N}=1$ supergravity to extract the characteristic data of the theory. Sections \ref{gaugeKin} and \ref{kaehlerPot} are devoted to finding the gauge kinetic functions, the bosonic parts of the chiral multiplets and the K\"ahler potential of the theory, respectively. The Green-Schwarz mechanism induces the gauging of an axionic Peccei-Quinn shift symmetry and the resulting D-term potential is discussed in section \ref{superpot}. It is also shown that the dualization of the auxiliary spacetime 3-forms introduced in the democratic formulation 
leads to a superpotential, similar in form to the flux superpotential due to RR 1- and 3-form fluxes, that depends on the dualisation scalars and forces them to be set to zero.

\subsection{Eliminating the additional degrees of freedom}
\label{dualization}
\noindent Having determined the entire pseudo-action of the theory we are now in a position to integrate out the redundant fields and obtain a true action for the physical fields. This is accomplished by introducing Lagrangian multiplier terms in such a way that the duality relations of equation (\ref{dualityG}) can be obtained from the equations of motion of the new action. Then the auxiliary fields in this new action may be integrated out and eliminated using their equations of motion.\\
The first step is to translate the duality conditions (\ref{dualityG}) of the ten-dimensional fields into corresponding equations of the 4D fields. To do this, one needs the identity
\begin{equation}
\ast_{\textbf{R}^{1,3}\times Y} (\omega\wedge\eta)= (-1)^{q(4-p)}(\ast_4\omega)\wedge(\ast_Y\eta)
\end{equation}
for a p-form $\omega$ on $\textbf{R}^{1,3}$ and a q-form $\eta$ on $Y$, which follows from the block-diagonal form of the metric. Using this together with the expansions of eq. (\ref{fieldExp}) one finds in the 4d Einstein frame
\begin{eqnarray}
\dd d_{(3)c}-\KK_{abc} \dd \BB^a\wedge \tilde{c}^b_{(3)}&=&0\label{dualD},\\
\dd \tilde{c}^a_{(3)}-\dd \BB^a\wedge c_{(3)}&=&0,\\
\dd c_{(3)}&=&0\label{dualC},\\
\dd \rho_{(2)\hat{K}}&=&\frac{6}{\KK}e^{\frac{\phi}{2}}A_{\hat{K}\hat{L}}\ast_4 \dd \xi^{\hat{L}}\label{dualRhoXi},\\
\dd U_{\alpha}-\KK_{a\alpha\beta}\dd \BB^a\wedge V^{\beta}&=&-\frac{2\KK}{3}e^{\frac{\phi}{2}}G_{\alpha\beta}\ast_4 \dd V^{\beta}.
\label{dualUV}\end{eqnarray}
The abbreviations used were defined in section \ref{section_Bulk}. Note that due to the fact that the background gauge fluxes are constant we have $\dd b^a=\dd \BB^a$, and the combinations of fields above are exactly the combinations appearing in the bulk and Chern-Simons actions.
\\Let us consider first the constraints involving the three-form fields in equations (\ref{dualD}) to (\ref{dualC}). Note that the three equations may be inserted into one another so that an equivalent system of constraints which is easier to implement is
\begin{eqnarray}
\dd \left( d_{(3)a} -\KK_{abc}\BB^b\tilde{c}^c_{(3)}\right)&=&0,\nn \\
\dd \left(\tilde{c}^a_{(3)}-\BB^a\wedge c_{(3)}\right)&=&0,\nn\\
\dd c_{(3)}&=&0.
\label{dualize3form}\end{eqnarray}
We will now exemplify the dualization for the field $d_{(3)a}$ but the results immediately carry over to the other two 3-form fields. The constraint has the form $D_{(4)a}:=\dd d_{(3)a} -\KK_{abc}\dd \BB^b\wedge \tilde{c}^c_{(3)}=\KK_{abc}\BB^b\dd\tilde{c}^c_{(3)}$ and the part of the action containing the relevant fields may be written as
\begin{equation}S=S_{rest}+\int \left[g^{ab} D_{(4)a}\wedge\ast_4 D_{(4)b} + h^aD_{(4)a}\right].\label{actionD}\end{equation}
Here $g^{ab},\ h^a$ denote functions containing constants as well as possibly the scalars of the theory. 
We now add a total spacetime derivative term to the action as a Lagrangian multiplier. This gives the constraints as the equation of motion of $\nu^a$, the constant dual to the three-form in $D_{(4)a}$ (compare \cite{louisMicu} and{ \cite{jockersD7}),
\begin{equation}S=S_{rest}+\int \left[g^{ab} D_{(4)a}\wedge\ast_4 D_{(4)b} + h^aD_{(4)a}+\nu^a D_{(4)a}-\nu^a\KK_{abc}\BB^b\dd\tilde{c}^c_{(3)}\right].\end{equation}
In this new action the constraints of (\ref{dualize3form}) arise as equations of motion rather than having to be imposed afterwards so we may eliminate $D_{(4)a}$ from the action by inserting its equation of motion
\begin{equation} g^{ab}\ast_4 D_{(4)b} = -\frac{h^a +\nu^a}{2} \end{equation}
(note the fact that $\ast_4\ast_4 D_{(4)a}=-D_{(4)a}$ due to the Lorentzian signature of the metric).\\
In the resulting action we see that $\nu^a$ plays the role of a cosmological constant and a potential for the scalar fields in $g^{ab},\ h^a$ is induced,
\begin{equation}S=S_{rest}+\int \frac{1}{4}g_{ab}(h^a+\nu^a)(h^b+\nu^b)\ast_4 1-\nu^a\KK_{abc}\BB^b\dd\tilde{c}^c_{(3)}.\label{actionDfinal}\end{equation}
Here $g_{ab}$ denotes the inverse matrix to $g^{ab}$.\\
There is another way to see that this action gives the correct dual description to the original action. The equation of motion of the three-form field $\dd d_{(3)a}$ resulting from the original action (\ref{actionD}) is
\begin{equation} \dd( 2g^{ab}\ast_4 (\dd d_{(3)b} -\KK_{bcd}\dd \BB^c\wedge \tilde{c}^d_{(3)})+h^a)=0.\end{equation}
This is solved by introducing a constant $\nu^a$ and setting
\begin{equation}g^{ab}\ast_4 (\dd d_{(3)b} -\KK_{bcd}\dd \BB^c\wedge \tilde{c}^d_{(3)})+\frac{h^a}{2}=-\frac{\nu^a}{2}.\end{equation}
Inserting the dualization condition we obtain the equation $\frac{1}{2}g_{ab}(h^b+\nu^b)=\KK_{abc}\BB^b\dd\tilde{c}^c_{(3)}$, which indeed also arises as the equation of motion of the action (\ref{actionDfinal}).\\
\indent The next step is to eliminate the spacetime two-form $\rho$ in favor of its dual scalar, $\xi$. This procedure is slightly more involved than the previous case of the three-forms. The problem lies in the fact that the duality relation of eq. (\ref{dualRhoXi}) is not compatible with the equation of motion of $\rho$. The part of the action containing $\rho$ and $\xi$ has the form
\begin{eqnarray}
S&=&S_{rest}-\frac{1}{2\kappa_4^2}\int \left[\frac{\KK}{24}e^{-\frac{\phi}{2}}A^{\hat{K}\hat{L}}H_{(3)\hat{K}}\wedge \ast_4 H_{(3)\hat{L}}+J^{\hat{K}}\wedge\dd\rho_{(2)\hat{K}}\right.\nn \\ &&\left.
+\frac{\KK}{24}e^{\frac{\phi}{2}}A^{\hat{K}\hat{L}}W_{(1)\hat{K}}\wedge\ast_4 W_{(1)\hat{L}}
+\xi^{\hat{K}}\dd\tilde{J}_{\hat{K}}+\frac{1}{2}\dd \xi^{\hat{K}}\wedge \dd \rho_{(2)\hat{K}}\right],
\label{actionRho}\end{eqnarray}
where $J$ and $\tilde{J}$ are one- and three-form source terms respectively, $H_{(3)\hat{K}}=\dd \rho_{(2)\hat{K}}$ is the field strength of the two-form and
\begin{equation}W_{(1)\hat{K}}=\frac{6}{\KK}A_{\hat{K}\hat{L}}\dd\xi^{\hat{L}}.\label{defW}\end{equation}
We have added the Lagrangian multiplier term $\frac{1}{2}\dd \xi^{\hat{K}}\wedge \dd \rho_{(2)\hat{K}}$ to the action to implement the duality constraint. The equation of motion for $\rho$ from this action is
\begin{equation}\dd\left(\frac{\KK}{12}e^{-\frac{\phi}{2}}A^{\hat{K}\hat{L}}\ast_4H_{(3)\hat{L}} -J^{\hat{K}}\right)=0,\end{equation}
which is incompatible with the duality condition $\ast_4 e^{-\frac{\phi}{2}}H_{(3)\hat{K}}=W_{(1)\hat{K}}$ with $W_{(1)\hat{K}}$ defined as in (\ref{defW}).\\
\indent The reason for this incompatibility lies in the fact that in the original ten-dimensional theory the duality relations would be implemented using an auxiliary field with the PST formalism, which was dropped for the purpose of compactifying the theory \cite{dallAgata, PST}. To restore consistency, we need to shift the definition of the field strength $W_{(1)\hat{K}}$ in the dualization condition to
\begin{equation}W_{(1)\hat{K}}=\frac{6}{\KK}A_{\hat{K}\hat{L}}\dd\xi^{\hat{L}}+\frac{12}{\KK}A_{\hat{K}\hat{L}} J^{\hat{L}}.\end{equation}
In a similar manner the source term for $\xi$ forces us to alter the Bianchi identity of $H_{(3)\hat{K}}=\dd\rho_{(2)\hat{K}}$, as the equation of motion of $\xi$ is 
\begin{equation}\dd\left(-\frac{1}{2}e^{\frac{\phi}{2}}\ast_4 W_{(1)\hat{K}}+\tilde{J}_{\hat{K}}\right)=0.\end{equation}
Hence we shift the 3-form field strength to
\begin{equation}H_{(3)\hat{K}}\rightarrow \dd\rho_{(2)\hat{K}}+2\tilde{J}_{\hat{K}}.\end{equation}
We are now left with the action
\begin{eqnarray}
S&=&S_{rest}-\frac{1}{2\kappa_4^2}\int \left[\frac{\KK}{24}e^{-\frac{\phi}{2}}A^{\hat{K}\hat{L}}H_{(3)\hat{K}}\wedge \ast_4 H_{(3)\hat{L}}+\frac{\KK}{12}A^{\hat{K}\hat{L}}W_{(1)\hat{K}}\wedge H_{(3)\hat{L}}\right.\nn \\ &&\left.
+\frac{\KK}{24}e^{\frac{\phi}{2}}A^{\hat{K}\hat{L}}W_{(1)\hat{K}}\wedge\ast_4 W_{(1)\hat{L}}
 -\frac{\KK}{6}A^{\hat{K}\hat{L}}W_{(1)\hat{K}}\wedge\tilde{J}_{\hat{L}}
+\xi^{\hat{K}}\dd\tilde{J}_{\hat{K}}\right] .
\end{eqnarray}
We see that we now have a consistent action in which the duality condition arises as the equation of motion of $H_{(3)\hat{K}}$ if we treat it as an independent field. We may now use this equation of motion to eliminate $H$ and finally obtain the action of the physical degrees of freedom,
\begin{eqnarray}
S&=&S_{rest}-\frac{1}{2\kappa_4^2}\int \left[
\frac{3}{\KK}e^{\frac{\phi}{2}}A_{\hat{K}\hat{L}}\dd\xi^{\hat{K}}\wedge\ast_4 \dd\xi^{\hat{L}} +\frac{12}{\KK}e^{\frac{\phi}{2}}A_{\hat{K}\hat{L}}\dd\xi^{\hat{K}}\wedge\ast_4 J^{\hat{L}}\right.\nn \\&&\left.
+2\xi^{\hat{K}}\dd\tilde{J}_{\hat{K}}+\frac{12}{\KK}e^{\frac{\phi}{2}}A_{\hat{K}\hat{L}}J^{\hat{K}}\wedge\ast_4 J^{\hat{L}} -2J^{\hat{K}}\wedge\tilde{J}_{\hat{K}}\right].
\label{actionRhoFinal}\end{eqnarray}

\indent Finally we need to eliminate the vector $U_{\alpha}$ in favour of its magnetic dual, $V^{\alpha}$. To this aim let us rewrite the dualization condition of eq. (\ref{dualUV}) as 
\begin{equation}e^{-\frac{\phi}{2}}\ast_4(\dd U_{\alpha}-\KK_{\alpha\beta a}\dd b^a\wedge V^{\beta})\equiv e^{-\frac{\phi}{2}}\ast_4\tilde{F}_{\alpha}=\frac{2\KK}{3}G_{\alpha\beta}\dd V^{\beta}=\frac{2\KK}{3}G_{\alpha\beta}F^{\beta}\label{dualUVnew}\end{equation}
(note that $\ast_4\ast_4 \tilde{F}_{\alpha}=-\tilde{F}_{\alpha}$).\\
Then the part of the action involving $U_{\alpha}$ and $V^{\alpha}$ and their field strengths $\tilde{F}_\alpha$, $F^{\alpha}$ may be written as
\begin{eqnarray}
S=&S_{rest}-\frac{1}{2\kappa_4^2}\int \left[\frac{3}{8\KK} e^{-\frac{\phi}{2}}G^{\alpha\beta}\tilde{F}_{\alpha}\wedge\ast_4\tilde{F}_{\beta}+\frac{\KK}{6}e^{\frac{\phi}{2}}G_{\alpha\beta}F^{\alpha}\wedge\ast_4 F^{\beta}\right.\nn\\&\left.
+\dd U_{\alpha}\wedge J^{\alpha}+V^{\alpha}\wedge \tilde{J}_{\alpha}\right],\label{actionUV}
\end{eqnarray}
where the source currents are given by
\begin{eqnarray}J^{\alpha}&=&-\frac{1}{4}\tilde{\mathcal{M}}^{I\alpha}\Phi_I F, \label{current1}\\
\tilde{J}_{\alpha}&=&\frac{1}{4}\left(\KK_{\alpha\beta a}\tilde{\mathcal{M}}^{I\beta} \BB^a \dd \Phi_I-\tilde{\mathcal{L}}^I_{\alpha}\dd a_I\right)\wedge F.\label{current2}\end{eqnarray}
We would now like to add a Lagrangian multiplier term $\frac{1}{4\kappa^2_4}\dd U_{\alpha}\wedge \dd V^{\alpha}$ to the action to implement the dualization constraint of (\ref{dualUV}). However, the equation of motion of $U_{\alpha}$ resulting from this action is
\begin{equation}\dd \left( \frac{3}{4\KK}e^{-\frac{\phi}{2}}G^{\alpha\beta}\ast_4\tilde{F}_{\beta}+J^{\alpha}-\frac{1}{2}\dd V^{\alpha}\right)=0,\label{eomU}\end{equation}
while the equation of motion of $V^{\alpha}$ together with the dualization condition gives
\begin{equation}\dd \left(\tilde{F}_{\alpha}+\dd U_{\alpha}\right)-\KK_{\alpha\beta a}\dd \BB^a\wedge F^{\beta}-2\tilde{J}_{\alpha}=0.\end{equation}
From (\ref{current1}, \ref{current2}) we can read off that $\tilde{J}_{\alpha}-\KK_{\alpha\beta a}\dd \BB^a\wedge J^{\beta}=\dd W_{\alpha}$ with
\begin{equation}W_{\alpha}=\frac{1}{4}\left(\KK_{\alpha\beta a}\tilde{\mathcal{M}}^{\beta I}\Phi_I\BB^a -\tilde{\mathcal{L}}^I_{\alpha}a_I\right) F .\end {equation}
We see that to achieve consistency of the equations of motion with the duality constraint we must shift the definitions of the field strengths in the kinetic terms and the dualization condition according to
\begin{equation}
F^{\alpha}\rightarrow \dd V^{\alpha}-2J^{\alpha}, \quad\quad\quad 
\tilde{F}_{\alpha}\rightarrow \dd U_{\alpha}-\KK_{\alpha\beta a}\dd \BB^a\wedge V^{\beta}+2W_{\alpha} .
\end{equation}
Now that we have a pseudo-action whose equations of motion are consistent with the dualization condition we may integrate out the magnetic vector $U_{\alpha}$ and eliminate it from the action using its equation of motion. The resulting action is
\begin{eqnarray}
S&=&S_{rest}-\frac{1}{2\kappa_4^2}\int \left[\frac{\KK}{3}e^{\frac{\phi}{2}}G_{\alpha\beta}\left(\dd V^{\alpha}-2J^{\alpha}\right)\wedge\ast_4\left(\dd V^{\beta}-2J^{\beta}\right)\right.\nn\\&&\left.
 +\frac{1}{2}\KK_{\alpha\beta a} \BB^a \dd V^{\alpha}\wedge\dd V^{\beta}
+2W_{\alpha}\wedge\left(\dd V^{\alpha}- J^{\alpha}\right)\right].
\label{actionUVFinal}\end{eqnarray}

\indent Finally we are in the position to read off the values of the sources and constants introduced in the dualization procedure and sum the different contributions to obtain the full action of the bulk and brane system in the Einstein frame,
\begin{eqnarray}
S^E&=&-\frac{1}{2\kappa^2_4}\int \left[
R\ast 1+\frac{1}{2}\dd\phi\wedge\ast\dd\phi+2G_{KL}\dd q^K\wedge\ast \dd q^{L} +2(G_{ab} +\frac{9}{4}\frac{\mathcal{K}_a\KK_b}{\KK^2}) \dd v^a\wedge\ast \dd v^b
\right.\nn \\&&
 +2e^{-\phi}G_{ab}\dd b^a\wedge\ast\dd b^b 
+\frac{3}{4\KK}e^{-\frac{\phi}{4}}\mathcal{H}^{IJ}\left(\dd \Phi_I\wedge\ast\dd \Phi_J +M^L_IM^K_J\dd\Phi_L\wedge\ast\dd\Phi_K \right. \nn \\&& \left. 
-2 M^L_I\dd\Phi_L\wedge\ast\dd a_J +\dd a_I\wedge\ast \dd a_J\right)
+\frac{3}{\KK}e^{\frac{\phi}{2}}A_{\hat{K}\hat{L}}\left[\nabla\xi^{\hat{K}}+2J^{\hat{K}}\right] \wedge\ast \left[\nabla\xi^{\hat{L}}+2J^{\hat{L}}\right]
\nn \\&&
+\frac{\KK}{3}e^{\frac{\phi}{2}}G_{\alpha\beta}\dd V^{\alpha}\wedge\ast\dd V^{\beta}+\frac{1}{2}\KK_{\alpha\beta a} \BB^a \dd V^{\alpha}\wedge\dd V^{\beta} \nn \\&&
+\frac{\KK}{3}e^{\frac{\phi}{2}}G_{\alpha\beta}\tilde{\mathcal{M}}^{I\beta}\Phi_I\dd V^{\alpha}\wedge\ast F 
+\frac{1}{2}\left(\KK_{\alpha\beta a}\tilde{\mathcal{M}}^{\beta I}\Phi_I\BB^a -\tilde{\mathcal{L}}^I_{\alpha}a_I\right)\dd V^{\alpha}\wedge F \nn \\&&
+\frac{1}{8} e^{-\frac{\phi}{4}}V_{\Pi_+}F\wedge\ast F 
+\frac{\KK}{12}e^{\frac{\phi}{2}}G_{\alpha\beta}\tilde{\mathcal{M}}^{I\alpha}\tilde{\mathcal{M}}^{J\beta}\Phi_J\Phi_I F\wedge\ast F 
\nn \\ &&
 + \frac{1}{8}\left(\KK_{\alpha\beta a}\tilde{\mathcal{M}}^{\beta I}\Phi_I\BB^a-\tilde{\mathcal{L}}^I_{\alpha} a_I\right)\tilde{\mathcal{M}}^{\alpha J}\Phi_J F\wedge F 
 \left.-\frac{1}{8}l_s^3\pi_{0\hat{K}}\xi^{\hat{K}}F\wedge F + V_{sc}
\right].
\label{actionFull}
\end{eqnarray}
The source current $J^{\hat{K}}$ arising from the dualization of the two-form $\rho$ is given by
\begin{equation}J^{\hat{K}}=-\frac{1}{8}\mathcal{Q}^{IJ\hat{K}}\left[\Phi_I\dd(a_J-M_J^L\Phi_L)-(a_J-M_J^L\Phi_L)\dd\Phi_I\right].   \end{equation}
$V_{sc}$ is the potential for the scalar fields arising from the dualization of the 3-forms. Up to terms of second order in the brane fields it is given by
\begin{eqnarray}
V_{sc}&=& 
\frac{1}{4l_s^2}\Big(\frac{6}{\KK}\Big)^3\, e^{-\frac{\phi}{2}}\left(\frac{1}{2}\mathcal{M}^{Ia}\Phi_I\KK_{abc}\BB^b\BB^c-\mathcal{L}_a^I a_I\BB^a-\eta_a\BB^a+\lambda \right)^2\ast_4 1\nn \\&&
+\frac{1}{16l_s^2}\Big(\frac{6}{\KK}\Big)^3\, e^{\frac{\phi}{2}}G^{ab}\left(\mathcal{L}_a^I a_I-\KK_{acd}\mathcal{M}^{Ic}\BB^d\Phi_I-\KK_{acd}\nu^c\BB^d+\eta_a\right)\nn \\&&
\times\left(\mathcal{L}_b^I a_I-\KK_{bef}\mathcal{M}^{Je}\BB^f\Phi_J-\KK_{bef}\nu^e\BB^f+\eta_b\right)\ast_4 1\nn \\&&
+\frac{6}{l_s^2\KK}e^{\frac{3}{2}\phi}G_{ab}\left(\mathcal{M}^{Ia}\Phi_I+\nu^a\right)\left(\mathcal{M}^{Jb}\Phi_J+\nu^b\right) .
\label{scalarPot}\end{eqnarray}
In the above $\lambda,\ \eta_a$ and $\nu^a$ are constants arising, by the spelled-out procedure, from elimination of the original three-forms appearing in (\ref{dualize3form}),
\bea
\label{duals}
\lambda \leftrightarrow c_{(3)}, \quad\quad \eta_a \leftrightarrow \tilde c_{(3)}^a, \quad\quad \nu^a   \leftrightarrow d_{(3)_a}.
\eea
\indent Note that in the action above the kinetic terms of the axionic scalars $\xi^{\hat{K}}$ contain the gauge covariant derivatives
\begin{equation}\nabla \xi^{\hat{K}} = \dd\xi^{\hat{K}}+\frac{1}{2l_s}\pi_0^{\hat{K}}A\label{covDeriv}.\end{equation}
The appearance of this covariant derivative is a manifestation of the Green-Schwarz mechanism for anomaly cancellation (reviewed for example in \cite{blumenhagen4Dcomp, blumenhagenCvetic}). It arises after dualization from the Green-Schwarz term in the Chern-Simons action which describes a linear coupling of the $U(1)$ field strength to the spacetime two-form $\rho$. The bulk action has a set of shift symmetries
\begin{equation}\xi^{\hat{K}}\rightarrow \xi^{\hat{K}}+c^{\hat{K}},\end{equation}
which are gauged upon inclusion of a D6-brane, resulting in the covariant derivatives given above. As reviewed e.g. in \cite{blumenhagen4Dcomp}, the non-gauge-invariant part of the Lagrangian obtained by gauging a shift symmetry of a Ramond-Ramond scalar fields appearing in the gauge kinetic functions (as is the case for the $\xi^{\hat{K}}$) exhibits exactly the right gauge transformation behavior to cancel the chiral anomalies. Note that this gauging also renders the brane $U(1)$ massive in general, with the mass proportional to
\begin{equation}A_{\hat{K}\hat{L}}\pi_0^{\hat{K}}\pi_0^{\hat{L}}=\int_Y \pi_-\wedge\ast \pi_- . \end{equation}

\subsection{The effective action in supergravity form}
\noindent Any $\mathcal{N}=1$ supergravity theory involving a set of chiral matter multiplets and vector multiplets is completely specified by the K\"ahler potential, the gauge kinetic couplings and the superpotential \cite{WessBagger, Bilal}. These can already be unambiguously determined from the bosonic part of the action, which takes the form
\begin{eqnarray}
S&=&-\frac{1}{2\kappa_4^2}\int \left[R^{(4)}\ast 1 + 2K_{M\bar{N}}\dd M^M\wedge \ast \dd \bar{M}^{\bar{N}} + (\mathrm{Re} f)_{\hat{\alpha}\hat{\beta}}F^{\hat{\alpha}}\wedge \ast F^{\hat{\beta}} \right.\nn \\&&\left.
+ (\mathrm{Im} f)_{\hat{\alpha}\hat{\beta}}F^{\hat{\alpha}}\wedge F^{\hat{\beta}} + 2(V_F+V_D)\ast 1\right].
\label{sugraForm}\end{eqnarray}
$M^M$ labels the bosonic components of the chiral multiplets while $F^{\hat{\alpha}}$ are the field strengths of the vector fields. $K_{M\bar{N}}=\partial_{M}\partial_{\bar{N}}K$ are the components of a K\"ahler metric, while the D- and F-term potentials may be computed from the superpotential $W$, the gauge coupling matrix and the K\"ahler covariant derivative $\mathrm{D}_M W=\partial_M W + (\partial_M K) W$ as 
\begin{equation}V_F=e^K\left(K^{M\bar{N}}\mathrm{D}_M W\mathrm{D}_{\bar{N}} \bar{W}-3|W|^2\right),\qquad V_D= \frac{1}{2}(\mathrm{Re}f^{-1})^{\hat{\alpha}\hat{\beta}}\DD_{\hat{\alpha}}\DD_{\hat{\beta}}.\label{FtermDterm}\end{equation}
The gauge kinetic coupling matrix  $f_{\hat{\alpha}\hat{\beta}}$ is a holomorphic function of the complex scalars of the theory, while the superpotential is the bosonic part of a chiral superfield built out of the other superfields of the theory \cite{CremmerSuperYM}. The indices $\hat{\alpha}$ run over the values $0,1\ldots, h^{(1,1)}_+(Y)$, and $F^0\equiv F$. The $F^\alpha$, $\alpha \neq 0,$ therefore label the bulk $U(1)$ fields from the Ramond-Ramond sector and $F^0$ refers to the brane $U(1)$.

 Finally, the $\DD_{\hat{\alpha}}$ are the Killing potentials of the gauged isometries of the target manifold \cite{WessBagger}. They are defined (up to an additive Fayet-Iliopoulos constant) by the requirement that
\begin{equation}K_{N\bar{M}}\bar{X}_{\hat{\alpha}}^{\bar{M}}=i\partial_{N}\DD_{\hat{\alpha}}, \end{equation}
where $X_{\hat{\alpha}}$ is the Killing vector belonging to the gauged isometry.\\
\indent The first task is to identify the correct chiral coordinates $M^M$ in terms of which our action may be written in the form of (\ref{sugraForm}). In the case of a IIA orientifold without a brane it has been shown in \cite{grimmOrientifold} that one part of the chiral spectrum consists of the complexified K\"ahler moduli, while the other fields are a combination of the complex structure moduli, the dilaton, and the scalar fields $\xi^{\hat{K}}$ from the reduction of the RR 3-form. We will see that the complexified K\"ahler moduli remain good chiral coordinates even in the presence of a D6-brane (at least in the limit of small deviations from a supersymmetric background that we are considering), while the other coordinates are corrected by a term involving the brane fields. The chiral coordinates can be most easily read off from the gauge kinetic functions, so we turn to these first.

\subsubsection{The gauge kinetic functions, chiral coordinates and kinetic mixing}
\label{gaugeKin}
\noindent The gauge kinetic matrix $f_{\hat{\alpha}\hat{\beta}}$ can be immediately read off from the full action of eq. (\ref{actionFull}). Its components are given by
\begin{eqnarray}
f_{\alpha\beta}&=&\frac{\KK}{3}e^{\frac{\phi}{2}}G_{\alpha\beta} + \frac{i}{2}\KK_{\alpha\beta a}\BB^a, \nn\\
f_{\alpha 0}=f_{0\alpha}&=& \frac{\KK}{6}e^{\frac{\phi}{2}}G_{\alpha\beta}\tilde{\mathcal{M}}^{I\beta}\Phi_I
+  \frac{i}{4}\left(\KK_{\alpha\beta a}\tilde{\mathcal{M}}^{I\beta}\Phi_I\BB^a -\tilde{\mathcal{L}}^I_{\alpha} a_I\right), \nn \\
f_{00}&=&\frac{1}{8}e^{-\frac{\phi}{4}}V_{\Pi_+}+\frac{\KK}{12}e^{\frac{\phi}{2}} G_{\alpha\beta}\tilde{\mathcal{M}}^{I\alpha}\tilde{\mathcal{M}}^{J\beta}\Phi_I\Phi_J \nn \\&&
- \frac{i}{4}\left[l_s^3\pi_{0\hat{K}}\xi^{\hat{K}}-\frac{1}{2}\left(\KK_{\alpha\beta a}\tilde{\mathcal{M}}^{J\beta}\Phi_J\BB^a -\tilde{\mathcal{L}}^I_{\alpha} a_I\right)\tilde{\mathcal{M}}^{I\alpha}\Phi_I\right] .
\label{gaugeKinComps}\end{eqnarray}
Recall that the couplings $\tilde{\mathcal{M}}^{I\alpha}$ and  $\tilde{\mathcal{L}}^I_{\alpha}$ have been defined in eq. (\ref{coeffCS}).

Using eq. (\ref{Gab}) and the fact that $\KK_{\alpha}=0$ it is straightforward to rewrite the first line as
\begin{equation}f_{\alpha\beta}=\frac{i}{2}\KK_{\alpha\beta a}\left(\BB^a+ie^{\frac{\phi}{2}}v^a\right)\equiv \frac{i}{2}\KK_{\alpha\beta a}t^a.\end{equation}
We see that just as in the case of a IIA orientifold without branes treated in \cite{grimmOrientifold} the kinetic matrix of the bulk $U(1)$ gauge bosons is a linear function of the complexified K\"ahler moduli
\bea
\label{KahlerMod}
t^a = \BB^a+ie^{\frac{\phi}{2}}v^a.
\eea
To simplify the couplings of the bulk and brane field strengths we note that as they are the coefficients of terms of second order in derivatives we may replace $\BB, J\rightarrow \BB_0, J_0$ in the expression for $f_{0\alpha}$ above. Then using (\ref{relOmegas}), (\ref{calibration}) and (\ref{exps_IB}) we find 
\begin{equation}\KK_{\alpha\beta a}\tilde{\mathcal{M}}^{J\beta}\BB^a_0=\frac{1}{l_s^3}\int_{\Pi_-}\left[ \iota^*(i_{s^J}\omega_{\alpha})\wedge\iota^*\BB_0+\iota^*\omega_{\alpha}\wedge M^J_H A^H\right]=M^J_H\tilde{\mathcal{L}}^H_{\alpha}.\label{relML}\end{equation}
By exactly the same reasoning
\begin{equation}e^{\frac{\phi}{2}}v^a_0\KK_{\alpha\beta a}\tilde{\mathcal{M}}^{J\beta}=\tilde{\mathcal{L}}^J_{\alpha}\label{relML2}\end{equation}
and hence
\begin{equation}
f_{\alpha 0}= -\frac{1}{4}\tilde{\mathcal{L}}^I_{\alpha}\left[\Phi_I+i\left(a_I-M^J_I\Phi_J\right)\right].\label{fmixing}\end{equation}
This shows that as we had anticipated the deformation and Wilson line moduli pair up into a chiral field given by
\begin{equation}u_I :=\Phi_I+i\left(a_I-M^J_I\Phi_J\right). \label{OpenMod}\end{equation}
\indent Finally we come to the gauge coupling of the brane $U(1)$. First of all we can use the argumentation of eq. (\ref{relML}) to simplify the part of $f_{00}$ involving $\Phi_I$ and $a_I$. Inserting the calibration condition (\ref{calibration}) we can then write the volume of the cycle wrapped by the brane in terms of the periods of $\Omega$ to obtain

\begin{eqnarray}
f_{00}=&-\frac{1}{8}\tilde{\mathcal{L}}^I_{\alpha}\tilde{\mathcal{M}}^{J\alpha} (\Phi_I\Phi_J + i\Phi_I(a_J - M^L_J \Phi_L))-\frac{i}{8}l_s^3\pi_{0\hat{K}}\xi^{\hat{K}} \nn \\&
+\frac{1}{8}l_s^3e^{-\phi}e^{\frac{1}{2}(K_{CS}-K_{KS})}\pi_{0\hat{K}}e^{-i\theta}\mathcal{Z}^{\hat{K}}.
\end{eqnarray}
Following \cite{grimmOrientifold} we define the 4d dilaton $D$ and the scale invariant variable $l^{\hat{K}}$ by 
\begin{eqnarray}
e^D &=& e^{\frac{\phi}{4}}\sqrt{\frac{3}{4\KK}} \;, \nonumber \\
l^{\hat{K}}&=&e^{-\phi}e^{\frac{1}{2}(K_{CS}-K_{KS})}e^{-i\theta}\mathcal{Z}^{\hat{K}}=e^{-D}e^{\frac12 K_{CS}}e^{-i\theta}\mathcal{Z}^{\hat{K}}.
\label{deflK}\end{eqnarray}
Here we used the explicit form of $K_{KS}$ given in (\ref{K_KS}). Note that due to the orientifold constraint $e^{-i\theta}\mathcal{Z}^{\hat{K}}$ and hence also $l^{\hat{K}}$ are real fields. Then we see that 
\begin{equation}f_{00}=-\frac{i}{8}l_s^3\pi_{0\hat{K}}N^{\hat{K}} -\frac{1}{16}\tilde{\mathcal{L}}^I_{\alpha}\tilde{\mathcal{M}}^{J\alpha}u_I u_J 
\end{equation}
is a holomorphic function of the fields $u_I$ and the complex field $N^{\hat{K}}$ defined by
\begin{equation}
\label{ComplMod}
N^{\hat{K}}=\xi^{\hat{K}} + i \left[ l^{\hat{K}}-\frac{1}{2l_s^3}\delta^{\hat K \hat L}\frac{\pi_{0\hat{L}}}{\left\|\pi_0\right\|^2} \tilde{\mathcal{L}}^I_{\alpha}\tilde{\mathcal{M}}^{J\alpha}u_I\bar{u}_J \right].
\end{equation}
Here $\left\|\pi_0\right\|^2 \equiv \pi_{0\hat{K}}\delta^{\hat K \hat L}\pi_{0\hat L}$. Note that equation (\ref{relML2}) and the symmetry of $\KK_{\alpha\beta}$ imply that $\tilde{\mathcal{L}}^I_{\alpha}\tilde{\mathcal{M}}^{J\alpha}$ is symmetric under exchange of $I$ and $J$. \\
If we set the brane fields $u_I$ to zero we again recover the chiral field of ref. \cite{grimmOrientifold} (the conventions differ slightly as we are working with Einstein frame fields).

To summarise, our analysis of the gauge kinetic function allowed us to deduce the form of the chiral superfields describing the K\"ahler moduli $t^a$ (cf. (\ref{KahlerMod})), the open string moduli $u_I$ (cf. (\ref{OpenMod})) and the complex structure moduli $N^{\hat K}$ (cf. (\ref{ComplMod})).
In the limit of small fluctuations the inclusion of a D6-brane leads to a correction of the complex structure moduli but not of the K\"ahler moduli of the bulk. In the case of a D7-brane in type IIB considered in \cite{jockersD7} this situation is exactly reversed. This was to be expected as the two cases are related by mirror symmetry, which exchanges the roles of K\"ahler and complex structure moduli. Note that in this computation of the gauge kinetic function we have restricted ourselves to the tree-level contribution of the massless Kaluza-Klein modes. Taking into account also massive KK modes one obtains threshold corrections to the D6-brane gauge coupling \cite{Lust:2003ky}, which however is beyond the scope of this paper. For recent advances in the computation of these one-loop corrections for the case of fractional D6-branes in toroidal orbifolds see \cite{Gmeiner:2009fb}.

Before continuing with the analysis of the K\"ahler potential let us briefly comment on the mixing between the bulk and brane gauge fields encoded in (\ref{fmixing}). In deriving this result the $\Phi_I$ were defined as the \textit{fluctuations} around the vacuum position of the brane. In particular their vacuum expectation value vanishes by definition. In this analysis it is therefore not possible to establish the existence of kinetic mixing between bulk and brane $U(1)$s, the phenomenological implications of which have been recently considered in \cite{Dimopoulos}. However, the Wilson line moduli $a_I$ may in principle develop a nonzero VEV $\left\langle a_I \right\rangle \neq 0$. The resulting Lagrangian term $\mathcal{L} \supset \left\langle a_I \right\rangle F^{\alpha}\wedge F$ does not contribute to the equations of motion but can give rise to observable consequences in the presence of a topologically non-trivial gauge field background \cite{magMixing}. In the presence of such gauge instantons the would-be shift symmetry of the axion $a_I$ is broken non-perturbatively.
Furthermore note that the gauge kinetic function we found does imply an interaction between the brane and bulk $U(1)$s with the fluctuating fields encoded in $\Phi_I$ and the Wilson lines. Interactions of this type are in principle amenable to light-shining-through-wall experiments. For recent discussions of the phenomenology of hidden $U(1)$s in similar contexts see e.g. \cite{AbelGoodsell_U1s,Goodsell:2010ie,Williams:2011qb} and references therein.

\subsubsection{The K\"ahler potential}
\label{kaehlerPot}
\noindent Now that we have identified the correct chiral variables of the theory the next step is to determine the K\"ahler potential. To do this we first have to rewrite the part of the Lagrangian $\mathcal{L}_{sc}$ containing the kinetic terms of the scalar fields in terms of the complex combinations found in the previous section. 
A straightforward calculation shows that 
\begin{eqnarray}
\left(2G_{ab}+\frac{9}{2}\frac{\KK_a\KK_b}{\KK^2}\right)\dd v^a\wedge\ast\dd v^b+2e^{-\phi}G_{ab}\dd\BB^a\wedge\ast\dd\BB^b+\frac{1}{2}\dd\phi\wedge\ast\dd\phi\nn \\
=2e^{-\phi}G_{ab}\dd t^a\wedge\ast\dd\bar{t}^b+2\dd D\wedge\ast \dd D .
\end{eqnarray}
Using eq. (\ref{Gab}) it is easy to check that $e^{-\phi}G_{ab}$ is a K\"ahler metric arising from the K\"ahler potential \begin{equation}
K_{KS}=-\log (\frac{4}{3}e^{\frac{3}{2}\phi}\KK)= 2D-2\phi.
\end{equation} 

Next we turn to the kinetic terms of the complex structure deformations and the 4d dilaton. Using Kodaira's formula (\ref{Kodaira}) we obtain for the variable $l^{\hat{K}}$ introduced in (\ref{deflK})    
\begin{eqnarray}
\dd l^{\hat{K}}&=&-l^{\hat{K}}\dd D +\frac{1}{2}(\partial_{q^L}K_{CS})l^{\hat{K}}\dd q^L+e^{-D}e^{\frac{1}{2}K_{CS}}\partial_{q^L}(e^{-i\theta} \mathcal{Z}^{\hat{K}})\dd q^L\nn \\
&=&-l^{\hat{K}}\dd D +e^{-D}e^{\frac{1}{2}K_{CS}}e^{-i\theta}\chi_L^{\hat{K}}\dd q^L-\frac{1}{2}(\partial_{q^L}K_{CS})l^{\hat{K}}\dd q^L.
\end{eqnarray}
Now we use the identities (\ref{relANnew}), (\ref{relationsN}) and (\ref{relsChiZ}) of special K\"ahler geometry derived in Appendix \ref{specGeom} to simplify the contractions of the periods arising in
\begin{eqnarray}
4e^{2D}A_{\hat{K}\hat{L}}\dd l^{\hat{K}}\wedge\ast\dd l^{\hat{L}}&=&2\dd D\wedge\ast\dd D+2\dd D\wedge\ast\dd K_{CS} \nn\\&&
+\frac{1}{2}\dd K_{CS}\wedge\ast \dd K_{CS}+2G_{KL}\dd q^K\wedge\ast \dd q^L.
\label{kineticL}\end{eqnarray}
Note that the terms mixing the K\"ahler and complex structure moduli fields involving $\dd K_{CS}$ above are not reproduced in our dimensional reduction. The reason for this discrepancy presumably lies in the fact that we only worked to lowest order in the dimensional reduction of the curvature scalar of appendix \ref{reduxR}. In evaluating the Christoffel symbols all indices were raised using the metric of the background complex structure. For example, there actually holds to linear order in $\delta z^{K}$
\begin{equation}\delta g^{ij}=-g^{i\bar{i}}g^{j\bar{j}}\delta g_{\bar{i}\bar{j}}\end{equation}
and hence
\begin{equation}\Gamma_{\mu j}^i=-\frac{i}{2}\partial_{\mu}v^a (\omega_a)_j^{\ i}-\frac{1}{2}\delta z^L g^{i\bar{i}}g^{k\bar{k}}(b_L)_{\bar{i}\bar{k}}(\bar{b}_{\bar{K}})_{kj}\partial_{\mu} \bar{z}^{\bar{K}}\end{equation}
with $b_L$ as introduced in eq. (\ref{defb}) of appendix \ref{reduxR}.
This yields mixing terms $\propto \dd v^a\dd z^K$ as well as additional terms of the form $\propto \dd z^K \dd z^L$ in the curvature tensor which should lead to the terms $\dd D \wedge \dd K_{CS}$  and $\dd K_{CS}\wedge\dd K_{CS}$ of equation (\ref{kineticL}). In the sequel we will assume that this is the case so that the part of the Lagrangian including the kinetic terms of the scalar fields encoded in eq. (\ref{actionFull})  may be written as
\begin{eqnarray}
\mathcal{L}_{sc}&=&2e^{-\phi}G_{ab}\dd t^a\wedge\ast\dd\bar{t}^b
+4e^{2D}A_{\hat{K}\hat{L}}\dd l^{\hat{K}}\wedge\ast\dd l^{\hat{L}}\nn \\&&
+4e^{2D}A_{\hat{K}\hat{L}}\left[\nabla\xi^{\hat{K}}+\frac{i}{8}\mathcal{Q}^{IJ\hat{K}}\left[\bar{u}_I\dd u_J-u_I\dd \bar{u}_J\right]\right]\nn \\&&
\wedge\ast\left[\nabla\xi^{\hat{L}}+\frac{i}{8}\mathcal{Q}^{IJ\hat{L}}\left[\bar{u}_I\dd u_J-u_I\dd \bar{u}_J\right]\right]+\frac{3}{4\KK}e^{-\frac{\phi}{4}}\mathcal{H}^{IJ}\dd u_I\wedge\ast\dd \bar{u}_J .
\label{lsc}
\end{eqnarray}
\indent To proceed with the evaluation of the K\"ahler potential requires more information about the integrals in (\ref{coeffDBI}) and (\ref{coeffCS}). It turns out that to find a K\"ahler potential reproducing the kinetic terms in the action we need to assume that
\begin{equation}
\mathcal{Q}^{IJ\hat{K}}=\frac{4}{l_s^3}\delta^{\hat K \hat L}\frac{\pi_{0\hat{L}}}{\left\|\pi_0\right\|^2} \tilde{\mathcal{L}}^I_{\alpha}\tilde{\mathcal{M}}^{J\alpha}.
\end{equation}
One can then use the relation $J_i^j\Omega_{jkl}=i \Omega_{ikl}$, where $J^j_i=g^{jk}J_{ik}$ are the components of the complex structure on $Y$, as well as the definition of the $A^I$ in eq. (\ref{relAS}) and the calibration condition (\ref{calibration}) to show that
\begin{equation}\ast_{\Pi_+}A^I=ie^{-\frac{\phi}{4}}e^{\frac{1}{2}(K_{CS}-K_{KS})}e^{-i\theta}\iota^*(i_{s^I}\Omega) . \end{equation}
With the relations of special geometry of appendix \ref{specGeom} and the fact that $\int_{\Pi_+}\iota^*(i_{s^I}\alpha_{\hat{K}})\wedge A^J=0$ due to the negative parity of the integrand we obtain
\begin{eqnarray}
-2ie^{2D}\mathcal{F}_{\hat{K}\hat{L}}l^{\hat{K}}\mathcal{Q}^{IJ\hat{L}} &=&-2 e^{2D}A_{\hat{K}\hat{L}}l^{\hat{K}} \mathcal{Q}^{IJ\hat{L}}  \nn \\
&=&-ie^{2D}e^{-\phi}  e^{\frac{1}{2}(K_{CS}-K_{KS})}e^{-i\theta}\mathcal{F}_{\hat{L}}\mathcal{Q}^{IJ\hat{L}} \nn \\ 
&=& i e^{-\phi } e^{2D} e^{\frac{1}{2}(K_{CS}-K_{KS})}e^{-i\theta}  \frac{1}{l_s^3}\int_{\Pi_+}\iota^*(i_{s^I}\Omega)\wedge A^J \nn \\ 
&=& \frac{3e^{-\frac{\phi}{4}}}{4\KK}\mathcal{H}^{IJ}.
\end{eqnarray}
After inverting the definition (\ref{deflK}) of $l^{\hat{K}}$ using (\ref{relationsN}) to obtain 
\begin{equation}e^{-2D}=2il^{\hat{K}}\mathcal{F}_{\hat{K}\hat{L}}l^{\hat{L}}\end{equation}
a straightforward calculation using (\ref{relNF}) shows that \cite{grimmOrientifold}
\begin{eqnarray}
\partial_{l^{\hat{K}}} D &=& -2i e^{2D}\mathcal{F}_{\hat{K}\hat{L}} l^{\hat{L}}, \nn\\
\partial_{l^{\hat{K}}}\partial_{l^{\hat{L}}} D &=& 2e^{2D}A_{\hat{K}\hat{L}}.
\end{eqnarray}
It is now straightforward but tedious to use the chain rule and the various equalities presented above to check that the kinetic terms of eq. (\ref{lsc}) may be written as
\begin{eqnarray}
\mathcal{L}_{sc}&=&2\partial_{\bar{M}}\partial_{N}K \dd N\wedge\ast\dd\bar{M}\nn\\
&=& 2K_{\bar{t}^at^b}\dd t^a\wedge\ast\dd \bar{t}^b + 2K_{\bar{N}^{\hat{K}}N^{\hat{L}}}\nabla N^{\hat{L}}\wedge\ast\nabla \bar{N}^{\hat{K}} \nn \\&&
+4\mathrm{Re}(K_{\bar{N}^{\hat{K}}u_I}\dd u_I\wedge\ast\nabla\bar{N}^{\hat{K}})+2K_{\bar{u}_Ju_I}\dd u_I\wedge\ast\dd\bar{u}_J
\end{eqnarray}
with the K\"ahler potential given by
\begin{equation}K=-\log\left(\frac{4}{3}e^{\frac{3}{2}\phi}\KK\right)+4D.\end{equation}
Here $\nabla N^{\hat{K}}=\dd N^{\hat{K}}+\frac{1}{2l_s}\pi_0^{\hat{K}}A$ is the covariant derivative introduced in section \ref{dualization} and the chiral fields  $N^{\hat K}$ associated with the complex structure moduli were defined in eq. (\ref{ComplMod}).\\
\indent Written in this way the K\"ahler potential is identical in form to the K\"ahler potential of a IIA orientifold without branes treated in \cite{grimmOrientifold}. The additional kinetic terms of the brane moduli are nevertheless reproduced due to the changed Jacobian factors resulting from the shifts of the chiral fields. This is because $l^{\hat{K}}$ and hence $D$ now has to be considered as a function of the $N^{\hat{K}}$ and the $u_I$. The same observation holds for D5- or D7-branes in IIB orientifolds \cite{jockersD7, grimmD5}. In contrast to the case of the D7-brane in type IIB of ref. \cite{jockersD7} it is possible to obtain an explicit expression for the K\"ahler potential as a function of the chiral fields:
\begin{eqnarray}
K&=-2\log\left(\frac{1}{2i}\mathcal{F}_{\hat{K}\hat{L}}\left[N^{\hat{K}}-\bar{N}^{\hat{K}}+\frac{i}{4}\mathcal{Q}^{IJ\hat{K}} u_I\bar{u}_J\right] \times\left[N^{\hat{L}}-\bar{N}^{\hat{L}}+\frac{i}{4}\mathcal{Q}^{IJ\hat{L}} u_I\bar{u}_J \right]\right) \nn \\ & -\log\left(\frac{i}{6}\KK_{abc}(t^a-\bar{t}^a)(t^b-\bar{t}^b)(t^c-\bar{t}^c)\right).
\label{KaehlerPotFinal}
\end{eqnarray}
\indent As we have just shown the moduli space of the theory splits locally into a direct product of two factors, one parametrized by the complexified K\"ahler moduli and the other by the complex structure and D-brane moduli. This splitting is well known in the case without D-branes \cite{grimmOrientifold}. The fact that the splitting persists in our K\"ahler potential upon addition of a D6-brane and the corrections affect only the part of the moduli space describing the complex structure deformations is to be expected because we have worked in the limit of small $\delta t^a$ and $u_I$ in imposing the calibration conditions of equation (\ref{calibration}).

\subsubsection{The scalar potential}
\label{superpot}
\noindent Having determined the K\"ahler potential and the gauge kinetic functions of the sigma model target space our next task is to interpret the scalar potential. Let us first look at the D-term potential induced by the gauging of the Peccei-Quinn symmetries. The only charged scalars are the $N^{\hat{K}}$ defined in (\ref{ComplMod}), whose covariant derivative reads
\begin{equation}\nabla N^{\hat{K}}=\dd N^{\hat{K}}+\frac{1}{2l_s}\pi_0^{\hat{K}} A.\end{equation}
This immediately allows us to read off the corresponding Killing vector on the moduli space,
\begin{equation}X=\frac{1}{2l_s}\pi_0^{\hat{K}} \frac{\partial}{\partial N^{\hat{K}}}.\end{equation}
As no fields are charged under any of the other gauge fields $V^{\alpha}$ this is the only Killing vector. The Killing potential is defined as the solution of \cite{WessBagger}
\begin{equation}K_{M \bar{P}}\bar{X}^{\bar{P}}=i\partial_{M} \DD,\end{equation}
where $M,\ P$ run over all chiral fields in the theory. As $X^{N^{\hat{K}}}$ is constant we find that up to a constant the Killing potential is given by
\begin{equation}\DD=-i\bar{X}^{\bar{N}^{\hat{K}}}\partial_{\bar{N}^{\hat{K}}}K=-\frac{2i}{l_s}e^{2D}\pi_0^{\hat{K}}\mathcal{F}_{\hat{K}\hat{L}}l^{\hat{L}}.\end{equation}
Note that $\DD$ is real as $\mathcal{F}_{\hat{K}\hat{L}}$ is purely imaginary. Hence the D-term potential of (\ref{FtermDterm}) will be real. To find the contribution to the action given by equation (\ref{FtermDterm}) we need to find the (0,0)-component of the inverse of the real part of the gauge kinetic matrix 
\begin{equation}\mathrm{Re} f=\left(\begin{array}{cc}-\frac{1}{8}\tilde{\mathcal{L}}^I_\alpha \tilde{\mathcal{M}}^{J\alpha} \Phi_I\Phi_J+\frac{l_s^3}{8}\pi_{0\hat{K}}l^{\hat{K}} &
-\frac{1}{4}\tilde{\mathcal{L}}^I_{\alpha}\Phi_I\\\\ -\frac{1}{4}\tilde{\mathcal{L}}^I_{\alpha}\Phi_I &
-\frac{1}{2}\KK_{\alpha\beta a}e^{\frac{\phi}{2}}v^a\end{array}\right).\end{equation}
Letting $\KK^{\alpha\beta}$ denote the inverse matrix to $\KK_{\alpha\beta}$ and using eq. (\ref{relML2}) one can explicitly check that the inverse of this matrix is given by
\begin{equation}(\mathrm{Re}f)^{-1}=\left(\begin{array}{lcr}\frac{8}{l_s^3\pi_{0\hat{K}}l^{\hat{K}}} &&
-\frac{2}{l_s^3\pi_{0\hat{K}}l^{\hat{K}}}\tilde{\mathcal{M}}^{I\beta}\Phi_I\\\\
-\frac{2}{l_s^3\pi_{0\hat{K}}l^{\hat{K}}}\tilde{\mathcal{M}}^{I\alpha}\Phi_I&&
-2e^{-\frac{\phi}{2}}\KK^{\alpha\beta}+\frac{1}{l_s^3\pi_{0\hat{K}}l^{\hat{K}}}\tilde{\mathcal{M}}^{I\alpha}\tilde{\mathcal{M}}^{J\beta}\Phi_I\Phi_J \end{array}\right).\end{equation}
Plugging the above into (\ref{FtermDterm}) we obtain the Fayet-Iliopoulos D-term potential
\begin{equation}
V_D=\frac{9}{l_s^8\KK^2}e^{\frac{1}{2}(K_{CS}-K_{KS})}\frac{\left(\frac{1}{l_s^3}\int_{\Pi_0}\mathrm{Im}(e^{-i\theta}\Omega)\right)^2}{\frac{1}{l_s^3}\int_{\Pi_0}\mathrm{Re}(e^{-i\theta}\Omega)}.
\label{DtermPot}\end{equation}
This is the well-known form of the D-term potential of a D6-brane that has already been obtained in \cite{Cremades:2002te} (see also e.g. \cite{blumenhagen4Dcomp, LuestIntBranes}). It measures the deviation from the calibration condition which states that the cycle wrapped by the brane must be volume minimizing, i.e. $\iota^* e^{-i\theta}\Omega \propto \mathrm{dvol}|_3$ or in other words $\mathrm{Im}(e^{-i\theta}\Omega|_{\Pi_0})=0$. It is important to keep in mind that we have used this calibration in our derivation of the effective action. Thus the scalar potential (\ref{DtermPot}) is not reproduced directly by straightforward dimensional reduction, but only indirectly via the formalism of gauged supergravity applied here. \\

\indent The next step would be to recast the scalar potential $V_{sc}$ obtained in eq. (\ref{scalarPot}) by dimensional reduction in the form of an F-term potential as given in (\ref{FtermDterm}). However, this turns out to be problematic as $V_{sc}$ contains terms mixing the K\"ahler structure moduli with the brane moduli. In finding the chiral coordinates $u_I$ defined in (\ref{OpenMod}) involving the brane fields  and deriving the part of the K\"ahler potential depending on them, we have used the calibration condition $\iota^* \BB=0 = \iota^* J$. To be able to write $V_{sc}$ as a consistent F-term we therefore have to evaluate $V_{sc}$ with $\BB$, $J$ replaced by the supersymmetric reference values $\BB_0$, $J_0$ satisfying (\ref{relAS}) and (\ref{relML}). These are by definition the values for which the brane moduli are unobstructed, so that in this approximation the superpotential is independent of the brane moduli.\\
\indent To derive a possible F-term potential for the brane moduli by dimensional reduction we would need to find a formulation for the $\mathcal{N}=1$ moduli space that allows the brane and K\"ahler moduli to be treated together. This is possible in the context of relative cohomology groups \cite{WittenTFT, ChiralRingsCohomology, LercheMayr_SpecGeomOpenClosed}.\footnote{For recent advances in the computation of brane superpotentials using various techniques see e.g.  \cite{Alim:2009bx,Grimm:2009ef,Grimm:2009sy,Jockers:2009ti,Alim:2010za,Grimm:2010gk}.} However, as argued in ref. \cite{KatzKachruSuperpot}, the superpotential of the brane moduli may actually receive non-vanishing contributions only non-perturbatively and the classical superpotential vanishes identically. The non-perturbative contributions arise as disk instantons whose boundaries form non-trivial one-cycles on the special Lagrangian cycle $\Pi_0$ \cite{MayrLerche_MirrorSymm, SuperPotandMirrorSymm, KachruKatzMcGreevy_MirrorSymm, DbranesOnQuintic}.\\
\indent Let us now take a closer look at the scalar potential (\ref{scalarPotConstants}) evaluated for the F-flat configuration in which we performed our dimensional reduction. From what we said above it reduces to
\begin{eqnarray}
V_{sc}&=&
\frac{1}{16l_s^2}(\frac{6}{\KK})^3e^{\frac{\phi}{2}}G^{ab}(\eta_a-\KK_{acd}\nu^c\BB^d)(\eta_b-\KK_{bef}\nu^e\BB^f)\ast_4 1\nn \\&&
+\frac{1}{4l_s^2}(\frac{6}{\KK})^3e^{-\frac{\phi}{2}}(\lambda-\eta_a\BB^a)^2\ast_4 1+\frac{6}{l_s^2\KK}e^{\frac{3}{2}\phi}G_{ab}\nu^a\nu^b\ast_4 1
\label{scalarPotConstants}\end{eqnarray}
and depends on the K\"ahler structure moduli $t^a$ as well as the dualization constants $\lambda, \nu^a, \eta_a$  of eq. (\ref{duals}). 
First we stress again that the K\"ahler moduli should be viewed as fixed at the F-flat value consistent with the calibration condition and are thus not to be treated dynamically in this potential. On the other hand, if we re-interpret the dualization constants $\lambda, \nu^a, \eta_a$ as dynamical fields, the positive definite $V_{sc}$ is minimised at $\lambda=\eta_a=\nu^a=0$. 
Since the dualisation constants were introduced as extra parameters in the dualisation procedure it is reassuring that 
going ``on-shell'' effectively removes these unphysical degrees of freedom.

Finally we make the following amusing observation: The scalar potential (\ref{scalarPotConstants}) follows as the F-term potential derived from a formal superpotential $W = W(t^a)$ that is treated as a function of the K\"ahler moduli only.
To see this we recall from  ref. \cite{grimmOrientifold} that $K_Q=4D$ obeys a no-scale condition
\begin{equation}(K_Q)_{n^{\hat{K}}}K_Q^{n^{\hat{K}}\bar{n}^{\hat{L}}}(K_Q)_{\bar{n}^{\hat{L}}}=4,\label{noScale}\end{equation}
where the $n^{\hat{K}}$ are obtained from our chiral fields $N^{\hat{K}}$ by setting the brane fields to zero. As $K_Q$ depends on $N^{\hat{K}}$ and $u_I$ only through $n^{\hat{K}}$ and (\ref{noScale}) is invariant under a change of coordinates of the target space we find that the no-scale condition still holds in the new coordinate system. Therefore taking a superpotential dependent only on the complexified K\"ahler moduli $W=W(t^a)$ the resulting F-term has the form
\begin{equation}V_F=e^{K_{KS}}e^{4D}\left(|W|^2+K_{KS}^{t^a \bar{t}^b}\mathrm{D}_{t^a}W\mathrm{D}_{\bar{t}^b}\bar{W}\right).\label{VfKaehlerOnly}\end{equation}
It is straightforward to check that in terms of the formal inverse $K^{ab}$ of $K_{ab}$ the inverse metric on the space of K\"ahler structure deformations is given by $e^{\phi}G^{ab}=2e^{\phi}v^av^b-\frac{2}{3}e^{\phi}\KK\KK^{ab}$ and therefore there holds
\begin{equation}\KK_a G^{ab}\KK_b=\frac{4}{3}\KK^2.\end{equation}
With these identities one may check by direct calculation that the scalar potential of eq. (\ref{scalarPotConstants}) results from the superpotential (cf. \cite{grimmOrientifold})
\begin{equation}W=\frac{4}{l_s}\lambda + \frac{4}{l_s}\eta_a t^a +\frac{2}{l_s}\KK_{abc}\nu^a t^b t^c.\end{equation}
It may be written in terms of the (string frame) complexified K\"ahler form $J_c=(\BB^a+ie^{\frac{\phi}{2}}v^a)\omega_a=t^a\omega_a$ as
\begin{equation}W=\frac{4}{l_s}\lambda +\frac{4}{l_s^7}\eta_a\int_Y \tilde{\omega}^a\wedge J_c +\frac{2}{l_s^7}\nu^a\int_Y \omega_a\wedge J_c\wedge J_c.\end{equation}
The fact that we were indeed able to rewrite the scalar potential obtained by dimensional reduction as an F-term potential as required by the general form of $\mathcal{N}=1$ supergravity is a satisfying consistency check of our calculation. Note that the dualization of the auxiliary non-dynamical 3-forms has led to a superpotential that formally is the same expression that was derived in \cite{grimmOrientifold} by switching on fluxes for the field strengths of the RR 1- and 3-forms.

\subsubsection{Scalar potentials by dimensional reduction}

As stressed several times, in our calculation of the effective action we have restricted ourselves to the true moduli fields, i.e. fields describing deformations of the brane and the ambient space that respect $\mathcal{N}=1$ supersymmetry. These supersymmetry conditions for the embedding of the D6-brane were obtained by requiring that the supersymmetry transformation of the fermionic fields on the brane can be cancelled by a kappa-symmetry transformation of the world-volume theory. From the four-dimensional point of view, we would expect the supersymmetry conditions to arise as equations specifying the flat directions of an effective potential. This potential renders the non-supersymmetric deformations massive.\\
\indent An example of such a potential term is given by the D-term potential of equation (\ref{DtermPot}). The minima of this potential are 3-cycles calibrated by $e^{-i\theta}e^{\frac{1}{2}(K_{CS}-K_{KS})}\Omega$. This should of course be read as a constraint on the deformations of both the bulk complex structure and the 3-cycle, and the true moduli of the theory parametrize the flat directions of this potential.\\
\indent The second set of supersymmetry conditions of eq. (\ref{calibration}) require the cycle wrapped by the brane to be Lagrangian with respect to the bulk K\"ahler form and the gauge connection on it to be flat. In a similar manner to the D-term condition mentioned above, one expects these conditions to arise from a corresponding potential term. These conditions are F-term conditions arising from a holomorphic superpotential $W$ by the requirement $0=F_i=\frac{\partial W}{\partial \phi_i}$ \cite{DbranesOnQuintic}, as can be seen by looking directly at the supersymmetry transformations of the fermionic fields \cite{Martucci}.\\
\indent It is not hard to find a candidate superpotential reproducing the conditions mentioned above \cite{Martucci, Thomas}. 
Let us introduce the notation $\mathcal{F} = \frac{l_s^2}{2\pi}  f - \iota^* B^{(2)}$ with $f$ the a priori unconstrained gauge flux along the brane.
Choosing a fixed reference cycle $\Pi_0$ and a 4-chain $\Gamma$ with boundary $\partial \Gamma=\Pi-\Pi_0$, a possible superpotential is given up to an additive and a multiplicative constant (which clearly do not affect the resulting F-term condition) by
\begin{equation}W=\int_{\Gamma}(\tilde{\mathcal{F}}+i J)^2.\end{equation}
Here $\tilde{\mathcal{F}}$ is an extension of ${\mathcal F}$ to all of $\Gamma$ satisfying $\tilde{\mathcal{F}}|_{\Pi_0}=\mathcal{F}_0$ and $\tilde{\mathcal{F}}|_{\Pi}=\mathcal{F}$ while $J$ is the K\"ahler form of the bulk. This function does not depend on the choice of the chain within a certain homology class. Choosing homologically different chains results in the addition of periods to $W$, which do not affect the resulting F-term conditions. The variation of this superpotential as we go from the cycle $\Pi$ to an infinitesimally displaced cycle $\Pi'$ obtained by following the flow of a normal vector field $\nu$ for a distance $\epsilon$ is given by \cite{Thomas}
\begin{equation}\delta W=\epsilon \int_{\Pi}i_{\nu}(\mathcal{F}+i J)^2=2i\epsilon \int_{\Pi}i_{\nu} J\wedge(\mathcal{F}+i J).\end{equation}
Similarly one can easily calculate the result of varying the $U(1)$-connection according to $A\rightarrow A + a$ with a one-form $a$. The fact that $a$ and $\nu$ were arbitrary show that the critical points of the potential are given by Lagrangian cycles and flat gauge bundles on them, i.e.
\begin{equation}J|_{\Pi}=0=\mathcal{F}.\end{equation}
As pointed in \cite{Martucci}, this can be extended to branes wrapped on cycles of other odd dimensions in type IIA theory with the general form of the superpotential given by
\begin{equation}W=c \int_{\Gamma} e^{i J}e^{\tilde{\mathcal{F}}}.\label{superpotIIA}\end{equation}
$\Gamma$ is of course an even-dimensional chain linking the wrapped cycle to a fixed reference cycle, as discussed above. In the mirror symmetric type IIB theory, the BPS D-branes are even-dimensional and the corresponding superpotential is obtained by replacing $e^{i J}$ with $\Omega$ in eq. (\ref{superpotIIA}) \cite{Martucci, Thomas}.\\
\indent Note that while the functional of eq. (\ref{superpotIIA}) gives the correct critical points we have not been able to \emph{derive} it by dimensional reduction. This is also true for the D-term potential of eq. (\ref{DtermPot}), which we obtained from the Killing vector of a gauged target space isometry rather than by direct dimensional reduction. This is of course not surprising since in our evaluation of the brane action we have limited ourselves to the moduli fields respecting the supersymmetry condition. It would be interesting to perform a dimensional reduction while allowing general deformations of the brane and the gauge field configuration and to evaluate the resulting superpotential. This would give a direct check of the superpotential mentioned above, which is obtained from more general considerations, and would give an explicit realization of the moduli space of the theory as the zero potential locus in the general field space. However, this approach immediately faces the challenge that the space of infinitesimal deformations of a 3-cycle with associated $U(1)$ connection is isomorphic to $\Omega^1(\Pi;\mathbf{R})\otimes\Omega^1(\Pi;\mathbf{R})$ \cite{Thomas}, and is therefore infinite dimensional. This means that we have no means of describing these general deformations in terms of a finite number of effective four-dimensional fields.\\
\indent One might hope to at least obtain a potential for the K\"ahler and complex structure moduli by dimensional reduction given a fixed brane and gauge field configuration $\Pi_0$, $\mathcal{F}_0$. However, this requires being able to express the pullback of the metric in the Dirac-Born-Infeld action in terms of the pullbacks of the K\"ahler form and the holomorphic (3,0)-form $\Omega$. This is difficult to do in general due to the fact that the relationships between the components of the metric and for example the K\"ahler form can be given directly only in complex coordinates, which do not exist on the 3-dimensional cycle. This is different for example in the case of a D7-brane in a type IIB compactification, where the cycle wrapped by the brane is 4-dimensional and inherits complex coordinates from the Calabi-Yau bulk. In this case the DBI action can be directly evaluated in terms of the K\"ahler form, as is performed for example in the appendix of \cite{HaackLuest}. Of course, in explicit realizations of the theory where the coordinates of the bulk are known the pullback can also be evaluated directly. This is for example carried out for the case of a D6-brane in a $T6$ orbifold in \cite{Villadoro}.

\section{Summary and Conclusion}
\label{conclusion}
In this paper we have evaluated the low-energy effective action of a spacetime-filling D6-brane supersymmetrically embedded in a generic type IIA orientifold. To this end we have performed a Kaluza-Klein compactification of the ten-dimensional type IIA supergravity in the democratic formulation of ref. \cite{bergshoeffIIAsugra} as well as of the Dirac-Born-Infeld and Chern-Simons actions governing the dynamics of the D6-brane. The resulting action has been rewritten in the standard form of $\mathcal{N}=1$ supergravity so that the K\"ahler potential, the gauge kinetic coupling functions and the superpotential can be extracted. In this analysis we have worked in the limit of small deformations of the brane and the bulk K\"ahler structure moduli. This analysis could in principle be extended to higher orders in these deformations using the theory of relative cohomology \cite{LercheMayr_SpecGeomOpenClosed}.

As has been anticipated for example in ref. \cite{KachruKatzMcGreevy_MirrorSymm}, we have found that the moduli parametrizing the deformations of the brane and the Wilson line moduli pair up to form the bosonic part of a chiral superfield. The complex structure variables of the bulk theory obtained in \cite{grimmOrientifold} are corrected by terms involving the brane moduli. At least in the limit of small deformations mentioned above, the K\"ahler metric maintains the block-diagonal form observed in the case without a D6-brane.

To summarise our main findings, the K\"ahler coordinates of the target space of the four-dimensional theory are given by
\begin{eqnarray*}
t^a &=& b^a-\frac{l_s^2}{2\pi} f^a+ie^{\frac{\phi}{2}}v^a, \\
u_I &=& \Phi_I + i(a_I - M^J_I \Phi_J), \\
N^{\hat{K}} &=& \xi^{\hat{K}} + i \left[ l^{\hat{K}}-\frac{1}{l_s^3}\delta^{\hat K \hat L}\frac{\pi_{0\hat{L}}}{\left\|\pi_0\right\|^2} \tilde{\mathcal{L}}^I_{\alpha}\tilde{\mathcal{M}}^{J\alpha}u_I\bar{u}_J \right].
\end{eqnarray*}
Here $b^a$, $v^a$ and $\xi^{\hat{K}}$ are the moduli of the Kalb-Ramond field, the K\"ahler form and the Ramond-Ramond 3-form, while $l^{\hat{K}}$ encodes the complex structure deformations. The brane excitations are encoded in the deformation moduli $\Phi_I$, the Wilson line moduli are denoted by $a_I$ and the gauge flux quanta are parametrised by $f^a$. The K\"ahler potential can be given explicitly in terms of these coordinates as
\begin{eqnarray*}
K&=-2\log\left(\frac{1}{2i}\mathcal{F}_{\hat{K}\hat{L}}\left[N^{\hat{K}}-\bar{N}^{\hat{K}}+\frac{i}{4}\mathcal{Q}^{IJ\hat{K}} u_I\bar{u}_J\right] \times\left[N^{\hat{L}}-\bar{N}^{\hat{L}}+\frac{i}{4}\mathcal{Q}^{IJ\hat{L}} u_I\bar{u}_J \right]\right) \\ & -\log\left(\frac{i}{6}\KK_{abc}(t^a-\bar{t}^a)(t^b-\bar{t}^b)(t^c-\bar{t}^c)\right).
\end{eqnarray*}

The addition of the D6-brane to the theory gauges the Peccei-Quinn shift symmetry of the axionic scalars $\xi^{\hat{K}}$. We have evaluated the resulting D-term potential in eq. (\ref{DtermPot}) and confirmed that it enforces the calibration condition requiring the cycle wrapped by the brane to minimize the volume. In this note we have not turned on background fluxes. However, we have included non-dynamical 3-forms in the Kaluza-Klein reduction. Upon rewriting the action in terms of the constants dual to these 3-forms we have found that for non-zero constants a superpotential is induced which is formally identical to the superpotential found in ref. \cite{grimmOrientifold} by turning on background fluxes for the field strengths of the RR 1- and 3-forms. The dualization constants are fixed at zero on-shell by minimizing this potential. 

The gauge coupling function of the $U(1)$ gauge theory on the world-volume of the D6-brane is computed in section \ref{gaugeKin}. We find
\begin{equation}
\mathrm{Re} f = \frac{1}{8}e^{-\frac{\phi}{4}}V_{\Pi_+}+\frac{\KK}{12}e^{\frac{\phi}{2}} G_{\alpha\beta}\tilde{\mathcal{M}}^{I\alpha}\tilde{\mathcal{M}}^{J\beta}\Phi_I\Phi_J.
\end{equation}
Note that the $\Phi_I$  parametrize deformations away from the vacuum configuration of the brane and so by definition have vanishing vaccum expectation value in the F-flat configuration. This means that as expected the gauge coupling strength of the four-dimensional $U(1)$ gauge theory is controlled by $V_{\Pi_+}$, the volume of the brane cycle measured in the Einstein frame. We have also commented on kinetic mixing between bulk and brane gauge fields which has recently been shown to have interesting phenomenological implications.

As the necessary conditions for a supersymmetric embedding of the brane depend on the moduli of the bulk theory, it is clear that the brane moduli cannot generally be treated independently. Rather, one should consider a total moduli space parametrized by all of the moduli together. This problem has been tackled in the setting of type IIB orientifolds in \cite{LercheMayr_SpecGeomOpenClosed} using the concepts of relative cohomology and variation of mixed Hodge structure. While a thorough treatment is beyond the scope of this note, we remark that a similar approach should be pursued in the type IIA setting to consistently treat the combined $\mathcal{N}=1$ moduli space. We expect that such a treatment will correct the K\"ahler potential and that the block-diagonal form of the metric on the moduli space holds only locally.

\subsection*{Acknowledgements}

We acknowledge interesting discussions with R. Blumenhagen, T. Grimm, A. Hebecker, H. Jockers, D. L\"ust and F. Marchesano. 
We thank T. Grimm and D. Lopes for coordinating the submission of their related work \cite{Grimm:2011dx}. T.W. is grateful to the Kavli Institute for Theoretical Physics, Santa Barbara and the Max-Planck-Institut f\"ur Physik, Munich for hospitality during part of this work.
This project was supported in part by the SFB Transregio 33 "The Dark Universe" and the NSF under Grant No.PHY05-51164.

\appendix

\section{Reduction of the curvature scalar}
\label{reduxR}
\noindent The Einstein-Hilbert term of the 10-dimensional supergravity action in the string frame has the form
\begin{equation}S_R^{sf}=-\frac{1}{2\kappa_{10}^2}\int\dd^{10}x\sqrt{-\hat{g}}e^{-2\phi}\hat{R}.\end{equation}
Using the general Weyl-rescaling formula in D dimension
\begin{equation}g_{mn}\rightarrow C^{-2}g_{mn} \Rightarrow \sqrt{-g}C^{D-2}R\rightarrow \sqrt{-g}(R+(D-1)(D-2)(\partial_{m}(\log C))^2)\label{WeylResc}\end{equation}
with $C=e^{-\frac{\phi}{4}}$ and $D=10$ we obtain this action in the Einstein-frame where the Einstein-Hilbert term is canonically normalized.
\begin{equation}
S_R=-\frac{1}{2\kappa_{10}^2}\int\dd^{10}x \sqrt{-\hat{g}}\left(\hat{R}+\frac{9}{2}\partial_{\mu}\phi\partial^{\mu}\phi\right)\label{RactionEF}\end{equation}
The term involving the dilaton $\phi$ can be combined with the (Weyl-rescaled) kinetic dilaton term from the supergravity action (\ref{sugraAct10D}), we will leave it out from now on and focus only on the reduction of the curvature scalar.\\
Using equations (\ref{var_g_mixed}) and (\ref{var_g_pure}) it is straightforward to find the Christoffel symbols up to first order in the moduli fields (i.e. indices are raised using $\hat{g}|_0$). In the following we will use the Einstein frame K\"ahler moduli such that (\ref{var_g_mixed}) remains unchanged. They are related to the string frame fields by $e^{\frac{\phi}{2}}v^a_E = v^a_{sf}$. Note that eq. (\ref{var_g_pure}) still holds after rescaling the metric, as the same factors of $e^{\frac{\phi}{2}}$ appear on both sides of the equation when rewriting it in terms of the rescaled metric. From now on all indices will be raised and lowered using the rescaled metric. In complex coordinates $z^j=y^j+iy^{j+3};\quad j=1, 2, 3$ the nonvanishing Christoffel symbols are the purely 4-dimensional ones and
\begin{equation}\begin{array}{rclcrcl}\Gamma^{\mu}_{ij}&=&-\frac{1}{2}\partial^{\mu}\bar{z}^K (\bar{b}_K)_{ij}=\overline{\Gamma^{\mu}_{\bar{i}\bar{j}}}&\quad&
\Gamma^{\mu}_{i\bar{j}}&=&\frac{i}{2}\partial^{\mu}v^a (\omega_a)_{i\bar{j}}\\
\Gamma_{\mu j}^i&=&-\frac{i}{2}\partial_{\mu}v^a (\omega_a)_{j\bar{k}}g^{i\bar{k}}&\quad&
\Gamma_{\mu \bar{j}}^{\bar{i}}&=&-\frac{i}{2}\partial_{\mu}v^a (\omega_a)_{i\bar{j}}g^{i\bar{i}}\\
\Gamma_{\mu \bar{j}}^i&=&\frac{1}{2}\partial_{\mu}z^K(b_K)_{\bar{j}\bar{k}}g^{i\bar{k}}=\overline{\Gamma_{\mu j}^{\bar{i}}}\end{array} \label{defb} \end{equation}
In the above $(b_K)_{\bar{i}\bar{j}}=\frac{-1}{||\Omega ||^2}\bar{\Omega}_{\bar{i}}^{\;kl}(\chi_{K})_{kl\bar{j}}$. Due to the fact that all Christoffel symbols with one internal and two external indices vanish the entries of the Riemann curvature tensor with only external indices agree with the 4-dimensional curvature tensor: $\hat{R}^{\mu}_{\nu\rho\tau}=(R^{(4)})^{\mu}_{\nu\rho\tau}$. However this is not true for the part involving internal indices, where we get
\begin{equation}\hat{R}^{m}_{npr}=(R^{(Y)})^m_{npr}+\Gamma^m_{\mu p}\Gamma^{\mu}_{nr}-\Gamma^m_{\mu r}\Gamma^{\mu}_{np}.\end{equation}
Using the symmetries of the Riemann tensor and the metric as well as the fact that $Y$ is Ricci-flat we obtain
\begin{equation}\hat{R}=R^{(4)}+4g^{\mu\nu}\hat{R}^i_{\mu i\nu}+g^{mn}(\Gamma^r_{\mu r}\Gamma^{\mu}_{mn}-\Gamma^r_{\mu n}\Gamma^{\mu}_{mr}).\end{equation}
Explicitly evaluating the above using the Christoffel symbols it is possible to show that up to a total 4-dimensional derivative (which is what we mean by ``$\cong$'') we obtain
\begin{eqnarray}
\sqrt{-\det\hat{g}}\hat{R}&\cong&\sqrt{-\det\hat{g}}\left[R^{(4)} -\frac{1}{2}(\partial_{\mu}z^K)(\partial^{\mu}\bar{z}^L)(b_K)_{\bar{i}\bar{j}}(\bar{b}_L)_{ij}g^{i\bar{i}}g^{j\bar{j}}\right.\nn\\&&\left.
 -  (\partial_{\mu} v^a)(\partial^{\mu} v^b) g^{i\bar{j}}g^{k\bar{l}}\left((\omega_a)_{i\bar{j}}(\omega_b)_{k\bar{l}}-\frac{1}{2}(\omega_a)_{i\bar{l}}(\omega_b)_{k\bar{j}}\right)
\right]\end{eqnarray}
\\
To write the metric on the K\"ahler structure moduli space in a concise manner we will need several identities. First of all we may use the fact that the harmonicity of $\omega_a$ implies that $\partial_m ((\omega_a)_{i\bar{j}}g^{i\bar{j}})=0$ and so $(\omega_a)_{i\bar{j}}g^{i\bar{j}}$ is a constant on Y. With this and the identities 
\begin{eqnarray}\epsilon_{ijk}\epsilon^{ljk}=2\delta^l_i, \qquad \epsilon_{ijk}\epsilon^{lmk}=2\delta^{lm}_{ij}=\delta^l_i\delta^m_j-\delta^l_j\delta^m_i\\
g_{i\bar{i}}g_{j\bar{j}}g_{k\bar{k}}\dd z^i\wedge \dd z^j \wedge \dd z^k=g_{i\bar{i}}g_{j\bar{j}}g_{k\bar{k}}\epsilon^{ijk}\dd^3z=\sqrt{g}\epsilon_{\bar{i}\bar{j}\bar{k}}\dd^3z
\label{epsIdentities}\end{eqnarray}
it is straightforward to show that $3i\frac{\KK_a}{\KK}=(\omega_a)_{i\bar{j}}g^{i\bar{j}}$. For an arbitrary (1,1)-form $\eta$ on $Y$ we then get
\begin{equation}\eta\wedge(-J\wedge\omega_a+\frac{3}{2}\frac{\KK_a}{\KK}J\wedge J)=\eta_{i\bar{j}}\omega_a^{i\bar{j}}\frac{J^3}{3!}\end{equation}
or (note that $\omega_a$ is real, so $\omega_a^{i\bar{j}}=\overline{\omega_a}^{i\bar{j}}$) \begin{equation}\ast\omega_a=-\left(J\wedge\omega_a-\frac{3}{2}\frac{\KK_a}{\KK}J\wedge J\right).\label{starOmega}\end{equation}
\indent The relation between the integration measures is $\dd^6y=-i\dd^6 z=-i\dd^3z\dd^3\bar{z}$, which can be used together with (\ref{epsIdentities}) to obtain
\begin{eqnarray}
\KK_{ab}&=&\frac{1}{l_s^6}\int_Y \omega_a\wedge\omega_b\wedge J\nn\\
&=&\frac{1}{l_s^6}\int_Y \dd^6y \sqrt{\det g^{(Y)}}g^{i\bar{j}}g^{k\bar{l}}\left((\omega_a)_{i\bar{j}}(\omega_b)_{k\bar{l}}-(\omega_a)_{i\bar{l}}(\omega_b)_{k\bar{j}}\right)\label{KabCoord}\end{eqnarray}
and
\begin{equation}\int_Y \dd^6y\sqrt{\det g^{(Y)}}(b_K)_{\bar{i}\bar{j}}(\bar{b}_L)^{\bar{i}\bar{j}}=4V_Y\frac{\int_Y \chi_K\wedge\bar{\chi}_L}{\int_Y \Omega\wedge\bar{\Omega}}.\end{equation}
Finally defining the metrics 
\begin{equation}G_{ab}=\frac{1}{4V_Y}\int_Y \omega_a\wedge\ast_Y\omega_b \stackrel{(\ref{starOmega})}{=} -\frac{3}{2}\left(\frac{\KK_{ab}}{\KK}-\frac{3}{2}\frac{\KK_a\KK_b}{\KK^2}\right)
\end{equation}
\begin{equation}
G_{KL}=-\frac{\int_Y \chi_K\wedge\bar{\chi_L}}{\int_Y \Omega\wedge\bar{\Omega}}
\end{equation}
we may perform the integrals over the internal space to get
\begin{eqnarray}
S_R&=&-\frac{1}{2\kappa_{10}^2}\int_{\mathbf{R}^{(1,3)}}\left[V_Y R^{(4)}\ast_4 1-\left(2V_YG_{ab}+l_s^6\KK_{ab}\right)\dd v^a\wedge\ast_4 \dd v^b \right.\nn \\&&\left.
+2V_Y G_{KL}\dd z^K\wedge\dd\bar{z}^L\right]
\end{eqnarray}
Note that in this expression the rescaling of the 4-dimensional metric of eq. (\ref{rescale4d}), which introduces further kinetic terms for the K\"ahler structure moduli via (\ref{WeylResc}), has not yet been carried out. \\
\indent As explained in section \ref{Ospectrum} in the orientifolded theory the $z^K$ are no longer arbitrary complex fields. The allowed metric fluctuations are parametrized by the real fields $q^K$ satisfying either $z^K=iq^K$ or just $z^K=q^K$. In either case $\dd q^K \partial_{q^K}=\dd z^K \partial_{z^K}$ restricted to $\mathcal{M}^{CS}_{\mathcal{N}=1}$. Kodaira's formula may be used to show that the metric on the complex structure moduli space is a K\"ahler metric
\begin{equation}G_{KL}=-\partial_{z^K}\partial_{\bar{z}^L} \log\left[\frac{i}{l_s^6}\int_Y \Omega\wedge\bar{\Omega}\right].\end{equation}
Hence we see that restricted to $\mathcal{M}^{CS}_{\mathcal{N}=1}$ we get $G_{KL}\dd z^K\wedge\dd\bar{z}^L=G_{KL} \dd q^K\wedge\dd q^L$.

\section{Pullback to the brane world-volume}
\label{pullbacks}

\noindent In the course of the reduction of the Dirac-Born-Infeld and Chern-Simons actions it is necessary to compute the pullbacks onto the brane world-volume of the bulk Ramond-Ramond-forms as well as the metric. These pullbacks encode the couplings of the bulk fields to the displacement moduli of the brane.\\
For small fluctuations around the rest position of the brane and mirror brane system, described by the canonically embedded world-volume
\begin{equation}\mathbf{R}^{1,3}\times\Pi_+=\mathcal{W}\stackrel{\iota}{\hookrightarrow}\mathcal{M}=\mathbf{R}^{1,3}\times Y,\end{equation}
the fluctuations may be written in terms of a section $\Phi$ of the normal bundle as described in section \ref{braneModuli}. The normal bundle of $\mathcal{W}$ in $\mathcal{M}$ can of course be identified with the normal bundle $N\Pi_+$ of $\Pi_+$ in $Y$, and we will therefore use the symbol $\exp$ to denote the exponential map on either of these spaces.\\
\indent As $Y$ and therefore also the closed subset $\Pi_+$ are compact there exists $\delta>0$ such that $\exp_p$ is is well-defined on $B_{\delta}(0)\subset N_p \Pi_+$ for all $p\in \Pi_+$. In fact, the tubular neighborhood theorem guarantees the existence of a neighborhood $V$ of the zero section in $N\Pi_+$ such that $B_{\delta}(0)\subset V\cap N_p \Pi_+ \;\forall p\in \Pi_+$ and $\exp$ is a diffeomorphism from $V$ onto an open neighborhood $U$ of $\Pi_+$ in $Y$. Here and in the following the metric on the normal space of $\Pi_+$ is of course induced by the metric of the ambient space $Y$. We assume the fluctuations to be sufficiently small such that the position of the fluctuating brane may be given as $\mathcal{W}_{\Phi}=\exp_{\Phi}(\mathcal{W})$, with 
\begin{equation}\exp_{\Phi}(p):=\exp_p(\frac{l_s}{2}\Phi(p)) \; \forall p\in \mathcal{W},\end{equation}
in particular we assume $||\frac{l_s}{2}\Phi||=||\frac{l_s}{2}\Phi_I s^I||<\delta$ on $\mathcal{W}$.\\
\indent We could now proceed to explicitly evaluate the pullback of a tensor field in this chosen tubular coordinate system. In the case of the metric it does indeed turn out to be necessary to go into local coordinates, however for the pullbacks of differential forms we will follow an equivalent but less index-laden method using the Lie derivative. To do this we note that every point in $U$ is connected by a unique geodesic to its basepoint on $\Pi_+$. We may therefore extend $\Phi$ to a vector field on all of $U$ by parallel transport along these geodesics. Let $\alpha$ be the local flow of this vector field, which for any point $(x,p)\in \mathbf{R}^{1,3}\times U$ obeys
\begin{equation}\dot{\alpha}(t,(x,p))=\frac{l_s}{2}\Phi_I(x)s^I(\alpha(t,(x,p))), \quad \alpha(0,(x,p))=(x,p).\end{equation}
The curves $\alpha(t,(x,p))$ for $(x,p)\in\mathcal{W}$ are exactly the geodesics $\exp_p(t\frac{l_s}{2}\Phi(x,p))$, as by definition geodesics are those curves with the property that they are parallel along themselves, i.e. that their tangent vector at a given point is equal to the parallel transport of the initial tangent vector along the curve. With $\alpha_t=\alpha(t,\cdot)$ we see that for a covariant tensor field $T$ on $Y$
\begin{equation}\exp_{\Phi}^* T= (\alpha_1 \circ \iota)^* T=\iota^* \alpha_1 ^*T.\end{equation}
On the other hand, the pullback along the flow of a vector field is given as the exponential of the Lie derivative along that field, so we obtain up to second order in derivatives
\begin{equation}\exp_{\Phi}^* T = \iota^* T + \iota^*\mathcal{L}_{\frac{l_s}{2}\Phi} T + \frac{1}{2}\iota^*\mathcal{L}_{\frac{l_s}{2}\Phi}\mathcal{L}_{\frac{l_s}{2}\Phi}T + \ldots \label{pullbackTensor}\end{equation}
\\Consider first the case of the metric. Using the coordinate expression for the Lie derivative given by
\begin{equation}(\mathcal{L}_vg)_{ij}=v^k g_{ij,k}+v^k_{,i}g_{kj}+v^k_{,j}g_{ik}\label{lie}\end{equation}
one obtains a rather lengthy expression involving numerous derivatives of $g$ as well as $\Phi$, which may be simplified by using the following facts.\\
Let $a, b $ denote directions along the brane while $m, n$ are normal to the brane. The basis of $TY$ at a point $(x^a,y^m\neq0)$ in the tubular neighborhood but not on the brane is defined by parallel transporting the basis of tangent vectors at the base point $(x^a,0)$ along the geodesic defining the tubular coordinates. As the Levi-Civita connection is metric and so scalar products of vectors are invariant under parallel transport all the terms with derivatives of either the metric or the components of $\Phi$ along normal directions vanish. Also, the orthogonality of the normal and tangent spaces immediately gives $\hat{g}_{ma}=0$. This simplifies the pullback of the metric to:
\begin{eqnarray}
P_{\Phi}\hat{g}^{sf}&:=&\exp_{\Phi}^* \hat{g}^{sf}\nn\\&
=&i^*\hat{g}^{sf} + \frac{l_s^2}{4}\hat{g}_{mn}^{sf}\partial_{a}\Phi^m\partial_{b}\Phi^n\dd x^{a}\otimes\dd x^{b}\nn\\
&=&\frac{6}{\KK}e^{\frac{\phi}{2}}\eta_{\mu\nu}^E\dd x^{\mu}\otimes\dd x^{\nu}+e^{\frac{\phi}{2}}g_{ij}^E\dd x^{i}\otimes\dd x^{j}\nn\\
&&+\frac{l_s^2}{4}e^{\frac{\phi}{2}}g^E(s^I,s^J)\partial_{\mu}\Phi_I\partial_{\nu}\Phi_J\dd x^{\mu}\otimes\dd x^{\nu}\nn\\
&&+\frac{l_s^2}{4}e^{\frac{\phi}{2}}\Phi_I\partial_{\mu}\Phi_J g^E_{mn}(s^I)^m_{,b}(s^J)^n\left(\dd x^{\mu}\otimes\dd x^b + \dd x^b\otimes\dd x^{\mu}\right)\nn\\&&
+\frac{l_s^2}{4}\Phi_I\Phi_J g^E_{mn}(s^I)^m_{,a}(s^J)^n_{,b}\dd x^a\otimes\dd x^b,
\end{eqnarray}
where we have moved to the Einstein frame by rescaling the metric as in equations (\ref{rescale}) and (\ref{rescale4d}).\\
\indent Apart from the metric all the other objects to be pulled back to the brane world-volume are (at least the internal part) closed differential forms. This is where the Lie derivative formalism is advantageous, because we may use Cartans formula for the Lie derivative of a differential form
\begin{equation}\mathcal{L}_v \omega = (\dd i_v + i_v \dd ) \omega\end{equation}
and see that it reduces to $\dd i_{\Phi}$ on closed forms. We are interested specifically in forms which may be written as $T=\omega\wedge\eta$, with $\omega$ a form on $\mathbf{R}^{1,3}$ and $\eta$ a closed form on $Y$. For such a form $\eta$ there holds
\begin{eqnarray}
\mathcal{L}_{\frac{l_s}{2}\Phi} \eta &=&\dd i_{\frac{l_s}{2}\Phi_I s^I} \eta \nn\\
& =& \frac{l_s}{2}\dd (\Phi_I i_{s^I}\eta)\nn \\
&=&\frac{l_s}{2}\dd \Phi_I\wedge i_{s^I}\eta + \frac{l_s}{2}\Phi_I\dd (i_{s^I}\eta),
\end{eqnarray}
Note that $\iota^* \dd (i_{s^I}\eta)$ is an exact form on $\Pi_+$, and any term that can be written as the wedge product of a form on 4d spacetime wedged with such an exact form will vanish when we integrate over the internal part of the brane world-volume. Hence such terms in the pullback may be dropped. The only case where we may not do this is for the Kalb-Ramond field in the DBI-action, where it appears in the determinant and not simply wedged with a closed form. Note as well that e.g. $\iota^* \dd (i_{\Phi}\eta)$ is not an exact form on $\Pi_+$ due to the $x$-dependence in $\Phi$. Keeping only the terms on the pullback which do not give a vanishing integral over $\Pi_+$ by the considerations above we finally obtain
\begin{eqnarray}
\exp_{\Phi}^* T&=&\omega\wedge\iota^*\eta + \frac{l_s}{2}\omega\wedge\dd \Phi_I \wedge\iota^*i_{s^I}\eta +\frac{l_s^2}{8}\omega\wedge\dd\Phi_I\wedge\dd\Phi_J\wedge\iota^*(i_{s^J}i_{s^I}\eta) \nn \\&& +\frac{l_s^2}{4}\Phi_I\omega\wedge\dd\Phi_J\wedge\iota^*(i_{s^J}\dd i_{s^I}\eta). 
\end{eqnarray}
We have also used the fact that the normal vector fields obey $\left[s^I,s^J\right]=0$ and hence $i_{s^J}\dd i_{s^I}\eta=i_{s^I}\dd i_{s^J}\eta$.\\
\indent For the expansion of the determinant we also give the pullback of the Kalb-Ramond two-form in normal coordinates obtained using the coordinate expression (\ref{lie}):
\begin{eqnarray}
(\exp_{\frac{l_s}{2}\Phi}^* \BB)&=&\frac{l_s^2}{4}\partial_{\mu}\Phi_I\partial_{\nu}\Phi_J \BB(s^I, s^J)\dd x^{\mu}\otimes\dd x^{\nu}
\nn\\&&
+\frac{l_s}{2}\partial_{\mu}\Phi_I (i_{s^I}\BB)_a\left(\dd x^a\otimes\dd x^{\mu}+\dd x^{\mu}\otimes\dd x^a\right)\nn \\&& +\frac{l_s^2}{4}\Phi_I(s^I)^m\partial_{\mu}\Phi_J (i_{s^J}\BB)_{a,m}\left(\dd x^{\mu}\otimes\dd x^a-\dd x^a\otimes\dd x^{\mu}\right)\nn \\&&
+\left(\frac{l_s}{2}\Phi_I(s^I)^m\BB_{ab,m}
+\frac{l_s^2}{8}\Phi_I(s^I)^m\Phi_J(s^J)^n\BB_{ab,mn}\right.\nn\\&&\left.
+\frac{l_s^2}{4}\Phi_I\Phi_J \BB_{mn}(s^I)^m_{,a}(s^J)^n_{,b}\right)\dd x^a\otimes \dd x^b.
\end{eqnarray}

\section{Special geometry of the $\mathcal{N}=2$ moduli space}
\label{specGeom}
\noindent It is well-known that the $\mathcal{N}=2$ complex structure moduli space $\mathcal{M}$ is a so-called special K\"ahler manifold, so that the entire information about the geometry of this space is encoded in a holomorphic prepotential \cite{FerraraSymplStructN=2, FerraraDualitySugraYM, VanProeyenSpecialKaehler, FreedSpecialKaehler}. In the following we will present a collection of identities arising from this special geometry which are necessary to rewrite the kinetic terms of the action in supergravity form.\\
A special K\"ahler manifold $\mathcal{M}$ is a Hodge-K\"ahler manifold with a line bundle $\mathcal{L}$, a symplectic vector bundle $\mathcal{H}$ over $\mathcal{M}$ and a holomorphic section $\Omega(z)$ of $\mathcal{L}$ such that in terms of the symplectic product $\left\langle .,.\right\rangle$ there holds
\begin{equation}K_{CS}=-\log(i\left\langle \Omega, \bar{\Omega}\right\rangle), \qquad \left\langle \Omega, \partial_{z^K} \Omega\right\rangle=0.\label{defK}\end{equation}
In our case $\mathcal{H}=H^3(Y)$ and the symplectic product is given by $\left\langle \alpha, \beta\right\rangle=\frac{1}{l_s^6}\int_Y \alpha\wedge\beta$, while $\Omega$ is the holomorphic $(3,0)$-form. As before the K\"ahler covariant derivatives of $\Omega$ are $\chi_K=\partial_{z^K}\Omega +\partial_{z^K} K_{CS} \Omega$, which can be written in terms of their periods as
\begin{equation}\Omega=\mathcal{Z}^{\hat{K}}\alpha_{\hat{K}}-\mathcal{F}_{\hat{K}}\beta^{\hat{K}};\qquad \chi_K=\chi_K^{\hat{L}}\alpha_{\hat{L}}-\chi_{K\hat{L}}\beta^{\hat{L}}.\end{equation}
It can be shown that the constraints imposed on the metric by the special geometry imply the existence of a complex symmetric matrix $\NN_{\hat{K}\hat{L}}$ satisfying 
\begin{equation}\mathcal{F}_{\hat{K}}=\NN_{\hat{K}\hat{L}}\mathcal{Z}^{\hat{L}};\qquad \chi_{K\hat{K}}=\bar{\NN}_{\hat{K}\hat{L}}\chi_K^{\bar{L}};\qquad (\mathrm{Im}\NN)_{\hat{K}\hat{L}}\mathcal{Z}^{\hat{K}}\bar{\mathcal{Z}}^{\hat{L}}=-\frac{1}{2}e^{-K_{CS}}.\label{relationsN}\end{equation}
Defining the $h^3(Y)\times h^3(Y)$ matrices 
\begin{equation}f_{\hat{K}}^{\hat{L}}=\left\{\begin{array}{c}\bar{\chi}_K^{\bar{L}}, \quad \hat{K}=K\neq 0\\ 
\mathcal{Z}^{\hat{L}}, \quad \hat{K}=0\end{array}\right\};\qquad h_{\hat{K}\hat{L}}=\left\{\begin{array}{c}\bar{\chi}_{K\bar{L}}, \quad \hat{K}=K\neq 0\\ 
\mathcal{F}_{\hat{L}}, \quad \hat{K}=0\end{array}\right\}\end{equation}
$\NN$ can be explicitly written as
\begin{equation}\NN_{\hat{K}\hat{L}}=h_{\hat{K}\hat{M}}(f^{-1})^{\hat{M}}_{\hat{L}},\label{defN}\end{equation}
while (\ref{relationsN}) may be used to rewrite (\ref{defK}) as
\begin{equation}G_{LK}=-2e^{K_{CS}}\bar{\chi}_K^{\hat{K}}\mathrm{Im}\NN_{\hat{K}\hat{L}}\chi_L^{\hat{L}};\qquad \mathcal{Z}^{\hat{K}}\mathrm{Im}\NN_{\hat{K}\hat{L}}\chi_K^{\hat{K}}=0.\label{relsChiZ}\end{equation}
The two sets of periods of $\Omega$ are not independent, in fact if we assume that the Jacobian matrix $\partial_{z^K}(\frac{\mathcal{Z}^L}{\mathcal{Z}^0})$ is invertible the condition $\left\langle \Omega,\partial_{z^K}\Omega\right\rangle=0$ can be used to show that the holomorphic prepotential $\mathcal{F}=\frac{1}{2}\mathcal{Z}^{\hat{K}}\mathcal{F}_{\hat{K}}$ satisfies
\begin{equation}\mathcal{F}_{\hat{K}}=\partial_{\mathcal{Z}^{\hat{K}}}\mathcal{F};\qquad \mathcal{F}_{\hat{K}\hat{L}}\mathcal{Z}^{\hat{L}}\equiv (\partial_{\mathcal{Z}^{\hat{K}}}\partial_{\mathcal{Z}^{\hat{L}}}\mathcal{F}) \mathcal{Z}^{\hat{L}}=\mathcal{F}_{\hat{K}}.\label{relFZ}\end{equation}
The prepotential is a homogenous function of degree two, hence $\mathcal{F}_{\hat{K}\hat{L}}$ does not depend on $\mathcal{Z}$.\\
We may also rewrite $\NN$ in terms of the periods and the (symmetric) matrix $\mathcal{F}_{\hat{K}\hat{L}}$ as
\begin{equation}
\NN_{\hat{K}\hat{L}}=\bar{\mathcal{F}}_{\hat{K}\hat{L}}+2i\frac{\mathrm{Im}\mathcal{F}_{\hat{K}\hat{M}}\mathcal{Z}^{\hat{M}}\mathrm{Im}\mathcal{F}_{\hat{L}\hat{N}}\mathcal{Z}^{\hat{N}}}{\mathcal{Z}^{\hat{M}}\mathrm{Im}\mathcal{F}_{\hat{M}\hat{N}}\mathcal{Z}^{\hat{N}}}.\label{relNF}\end{equation}
This equation is most easily checked by multiplying both sides with the (invertible) matrix $f^{\hat{K}}_{\hat{L}}$ and using the previous relations. Finally, one can use the fact that $\ast \Omega = -i \Omega$ to write the matrix $A_{\hat{K}\hat{L}}$ of eq. (\ref{defA_KL}) in terms of the periods. The result is \cite{grimmOrientifold, Suzuki}
\begin{equation}A_{\hat{K}\hat{L}}=-\mathrm{Im}\NN_{\hat{K}\hat{L}}-(\mathrm{Re}\NN(\mathrm{Im}\NN)^{-1}\mathrm{Re}\NN)_{\hat{K}\hat{L}}.\label{relAN}\end{equation}
\indent Let us now turn to the case of the restricted moduli space obtained after performing the orientifold projection. In this case (\ref{condF}) shows that $e^{-i\theta}\mathcal{Z}^{\hat{K}}$ is real while $e^{-i\theta}\mathcal{F}_{\hat{K}}$ is purely imaginary. It follows that $e^{-i\theta}\chi_{K}^{\hat{K}}$ is real while $e^{-i\theta}\chi_{K\hat{K}}$ is purely imaginary, such that $\NN$ turns out to be purely imaginary. Hence (\ref{relAN}) simplifies to
\begin{equation}A_{\hat{K}\hat{L}}=-\mathrm{Im}\NN_{\hat{K}\hat{L}}.\label{relANnew}\end{equation}
Similarly eq. (\ref{relFZ}) and the orientifold conditions (\ref{condF}) show that $\mathcal{F}_{\hat{K}\hat{L}}$ is purely imaginary. In the orientifold case the complex structure K\"ahler potential becomes a simple function of the prepotential
\begin{equation}e^{-K_{CS}}=4ie^{-2i\theta}\mathcal{F}. \label{relKF}\end{equation}


\newpage
\begin{thebibliography}{9}

\bibitem{LuestIntBranes}D. L\"ust, ``Intersecting brane worlds: A path to the standard model?'', Class.
Quant. Grav. 21 (2004) S1399-1424, \htmladdnormallink{arXiv:hep-th/0401156}{http://arxiv.org/abs/hep-th/0401156}

\bibitem{blumenhagenCvetic}R. Blumenhagen, M. Cveti$\check{\textnormal{c}}$, P. Langacker and G. Shiu, "Toward Realistic Intersecting D-Brane Models", \htmladdnormallink{arXiv:hep-th/0502005v2}{http://arxiv.org/abs/hep-th/0502005v2}

\bibitem{blumenhagen4Dcomp}R. Blumenhagen, B. K\"ors, D. L\"ust and S. Stieberger, "Four-dimensional String Compactifications with D-Branes, Orientifolds and Fluxes", \htmladdnormallink{arXiv:hep-th/0610327v3}{http://arxiv.org/abs/hep-th/0610327v3}

\bibitem{DbraneModelMarchesano}F. G. Marchesano, ``Progress in D-brane model building'', \htmladdnormallink{arXiv:hep-th/0702094v2}{http://arxiv.org/abs/hep-th/0702094v2}


\bibitem{Blumenhagen:2009qh}
  R.~Blumenhagen, M.~Cveti{\v c}, S.~Kachru and T.~Weigand,
  ``D-Brane Instantons in Type II Orientifolds,''
  Ann.\ Rev.\ Nucl.\ Part.\ Sci.\  {\bf 59} (2009) 269,
   \htmladdnormallink{arXiv:0902.3251 [hep-th]}{http://arxiv.org/abs/0902.325}
  


\bibitem{Cvetic:2011vz}
  M.~Cveti{\v c} and J.~Halverson,
 ``TASI Lectures: Particle Physics from Perturbative and Non-perturbative Effects in D-braneworlds,''
 \htmladdnormallink{arXiv:1101.2907 [hep-th]}{http://arxiv.org/abs/1101.2907}


\bibitem{GranaFlux}M. Gra$\tilde{\textnormal{n}}$a, "Flux compactifications in string theory: A comprehensive review",
Phys. Rept. 423 (2006) 91-158, \htmladdnormallink{arXiv:hep-th/0509003}{http://arxiv.org/abs/hep-th/0509003}

\bibitem{DouglasKachruFlux}M. R. Douglas and S. Kachru, "Flux compactification", \htmladdnormallink{arXiv:hep-th/0610102}{http://arxiv.org/abs/hep-th/0610102}

\bibitem{SoftSusyBreakingGranaGrimm}M. Gra$\tilde{\textnormal{n}}$a, T. W. Grimm, H. Jockers and J. Louis, "Soft supersymmetry breaking in Calabi-Yau orientifolds with D-branes and fluxes", \htmladdnormallink{arXiv:hep-th/0312232}{http://arxiv.org/abs/hep-th/0312232}

\bibitem{jockersD7}H. Jockers and J.Louis, "The effective action of D7-branes in $\mathcal{N}=1$ Calabi-Yau orientifolds", \htmladdnormallink{arXiv:hep-th/0409098v3}{http://arxiv.org/abs/hep-th/0409098v3}


\bibitem{Jockers:2005zy}
  H.~Jockers and J.~Louis,
  ``D-terms and F-terms from D7-brane fluxes,''  
  Nucl.\ Phys.\  B {\bf 718} (2005) 203,
   \htmladdnormallink{arXiv:hep-th/0502059}{http://arxiv.org/abs/hep-th/0502059}
   

\bibitem{Jockers:2008pe}
  H.~Jockers, M.~Soroush,
  ``Effective superpotentials for compact D5-brane Calabi-Yau geometries,''
  Commun.\ Math.\ Phys.\  {\bf 290 } (2009)  249-290,  \htmladdnormallink{arXiv:0808.0761 [hep-th]}{http://arxiv.org/abs/0808.0761}

  
  
   

\bibitem{grimmD5}T. Grimm, T.-W. Ha, A. Klemm and D. Klevers, "The D5-brane effective action and superpotential in $\mathcal{N}=1$ compactifications", \htmladdnormallink{arXiv:hep-th/0811.2996v3}{http://arxiv.org/abs/0811.2996v3}


\bibitem{grimmOrientifold}T. Grimm and J. Louis, "The effective action of type IIA orientifolds", \htmladdnormallink{arXiv:hep-th/0412277v2}{http://arxiv.org/abs/hep-th/0412277v2}



\bibitem{louisMicu}J. Louis and A. Micu, "Type II theories compactified on Calabi-Yau threefolds in the presence of background fluxes", Nucl. Phys. B \textbf{635} (2002) 395 ,  \htmladdnormallink{arXiv:hep-th/0202168}{http://arxiv.org/abs/hep-th/0202168}

\bibitem{bergshoeffIIAsugra}E. Bergshoeff, R. Kallosh, T. Ort�n, D. Roest and A. Van Proeyen, "New Formulations of D=10 Supersymmetry and D8=O8 Domain Walls", \htmladdnormallink{arXiv:hep-th/0103233v2}{http://arxiv.org/abs/hep-th/0103233v2}

\bibitem{LercheMayr_SpecGeomOpenClosed}W. Lerche, P. Mayr and N. Warner, "Holomorphic N = 1 special geometry of open-closed type II strings", \htmladdnormallink{arXiv:hep-th/0207259}{http://arxiv.org/abs/hep-th/0207259};
"N = 1 special geometry, mixed Hodge variations
and toric geometry", \htmladdnormallink{arXiv:hep-th/0208039}{http://arxiv.org/abs/hep-th/0208039}

\bibitem{orientifoldMirrorSymm_Vafa}B. Acharya, M. Aganagic, K. Hori and C. Vafa, "Orientifolds, mirror symmetry and superpotentials", \htmladdnormallink{arXiv:hep-th/0202208}{http://arxiv.org/abs/hep-th/0202208}

\bibitem{bandosDemocSugra}I. Bandos, A. Nurmagambetov and D. Sorokin, "Various Faces of Type IIA Supergravity", \htmladdnormallink{arXiv:hep-th/0307153v1}{http://arxiv.org/abs/hep-th/0307153v1}

\bibitem{moduliCandelas}P. Candelas and X. de la Ossa, "Moduli Space Of Calabi-Yau Manifolds", Nucl.
Phys. B \textbf{355} (1991) 455 



\bibitem{Bachas:2008jv}
  C.~Bachas, M.~Bianchi, R.~Blumenhagen, D.~L{\"u}st, T.~Weigand,
  ``Comments on Orientifolds without Vector Structure,''
  JHEP {\bf 0808 } (2008)  016, \htmladdnormallink{arXiv:0805.3696 [hep-th]}{http://arxiv.org/abs/0805.3696}
  


\bibitem{dallAgata}G. Dall' Agata, "Type IIB supergravity compactified on a Calabi-Yau manifold with H-fluxes", \htmladdnormallink{arXiv:hep-th/0107264}{http://arxiv.org/abs/hep-th/0107264}


\bibitem{gukovHaack}S. Gukov and M. Haack, "IIA string theory on Calabi-Yau fourfolds with background fluxes", Nucl. Phys. B \textbf{639} (2002) 95, \htmladdnormallink{arXiv:hep-th/0203267}{http://arxiv.org/abs/hep-th/0203267}


\bibitem{ferraraDimRed}S. Ferrara and S. Sabharwal, "Dimensional Reduction Of Type II Superstrings",
Class. Quant. Grav. \textbf{6} (1989) L77

\bibitem{sugraLagBodner}M. Bodner, A. C. Cadavid and S. Ferrara, "(2,2) Vacuum Configurations For Type
Iia Superstrings: N=2 Supergravity Lagrangians And Algebraic Geometry", Class.
Quant. Grav. \textbf{8} (1991) 789

\bibitem{branesAtAngles}M. Berkooz, M. R. Douglas and R. G. Leigh, "Branes intersecting at angles," Nucl.
Phys. B 480 (1996) 265 \htmladdnormallink{arXiv:hep-th/9606139}{http://arxiv.org/abs/hep-th/9606139}

\bibitem{5branesMembranes_BeckerStrom}K. Becker, M. Becker, and A. Strominger, ``Five-branes, membranes and
nonperturbative string theory'', Nucl. Phys. \textbf{B456} (1995) 130-152, \htmladdnormallink{arXiv:hep-th/9507158}{http://arxiv.org/abs/hep-th/9507158}

\bibitem{susy3cycles_Kachru}S. Kachru and J. McGreevy, ``Supersymmetric three-cycles and (super)symmetry
breaking'', Phys. Rev. D61 (2000) 026001, \htmladdnormallink{arXiv:hep-th/9908135}{http://arxiv.org/abs/hep-th/9908135}

\bibitem{susypbranes_Strominger}M. Marino, R. Minasian, G. W. Moore, and A. Strominger, ``Nonlinear
instantons from supersymmetric p-branes'', JHEP 01 (2000) 005, \htmladdnormallink{arXiv:hep-th/9911206}{http://arxiv.org/abs/hep-th/9911206}


\bibitem{Freed:1999vc} D.~S.~Freed and E.~Witten,``Anomalies in string theory with D-branes,''  \htmladdnormallink{arXiv:hep-th/9907189}{http://arxiv.org/abs/hep-th/9907189}

\bibitem{Bryant:1998wv}
  R.~L.~Bryant, E.~R.~Sharpe,
  ``D-branes and spin**c structures,''
  Phys.\ Lett.\  {\bf B450 } (1999)  353-357.
  \htmladdnormallink{arXiv:hep-th/9812084}{http://arxiv.org/abs/hep-th/9812084}

\bibitem{ChiralRingsCohomology}W. Lerche, C. Vafa and N. P. Warner, ``Chiral Rings In N=2 Superconformal Theories'', Nucl. Phys. B 324, 427 (1989)

\bibitem{WittenTFT}E. Witten, ``Mirror manifolds and topological field theory'', \htmladdnormallink{arXiv:hep-th/9112056}{http://arxiv.org/abs/hep-th/9112056}


\bibitem{McLean_Slag}R.C.McLean, ``Deformations of calibrated submanifolds'', Comm. Anal. Geom. 6, (1998), 705-747.

\bibitem{Hitchin_Slag} N.J. Hitchin, ``The moduli space of special Lagrangian
submanifolds'', \htmladdnormallink{arXiv:dg-ga/9711002v1}{http://arxiv.org/abs/dg-ga/9711002v1}

\bibitem{Salur_SlagDef}S. Salur, ``Deformations of Special Lagrangian Submanifolds'', Communications in Contemporary Mathematics Vol.2, No.3 (2000), 365-372

\bibitem{Salur_Fred}S. Salur, ``Deformations of special Lagrangian submanifolds;
An approach via Fredholm alternative'', Proceedings of 12th G\"okova Geometry-Topology Conference pp. 157 - 164, GokovaGT.org

\bibitem{GrossJoyce_CYGeom}M. W. Gross, D. Huybrechts, D. D. Joyce, ``Calabi-Yau manifolds and related geometries: lectures at a summer school in Nordfjordeid'', Springer Verlag, Heidelberg 2003


\bibitem{polchinskiOrientifold}E. G. Gimon and J. Polchinski, "Consistency Conditions for Orientifolds and DManifolds", Phys. Rev. D 54 (1996) 1667 , 
\htmladdnormallink{arXiv:hep-th/9601038}{http://arxiv.org/abs/hep-th/9601038}

\bibitem{MayrLerche_MirrorSymm}W. Lerche and P. Mayr, "On N = 1 mirror symmetry for open type II strings",
\htmladdnormallink{arXiv:hep-th/0111113}{http://arxiv.org/abs/hep-th/0111113}

\bibitem{Lerche_SpecGeom}W. Lerche, "Special geometry and mirror symmetry for open string
backgrounds with N = 1 supersymmetry", \htmladdnormallink{arXiv:hep-th/0312326}{http://arxiv.org/abs/hep-th/0312326}

\bibitem{TASIpolchinski}J. Polchinski, ``TASI lectures on D-Branes'', \htmladdnormallink{arXiv:hep-th/9611050v2}{http://arxiv.org/abs/hep-th/9611050v2}


\bibitem{polchinskiRRCharges}J. Polchinski, ``Dirichlet-branes and Ramond-Ramond charges'', Phys. Rev. Lett.
75 (1995) 4724-4727, \htmladdnormallink{arXiv:hep-th/9510017}{http://arxiv.org/abs/hep-th/9510017}


\bibitem{DilatonTadpoles}R. Rabadan and F. Zamora, ``Dilaton tadpoles and D-brane interactions in
compact spaces'', JHEP 12 (2002) 052, \htmladdnormallink{arXiv:hep-th/0207178}{http://arxiv.org/abs/hep-th/0207178}

\bibitem{tadpolesVacuumredef}E. Dudas, G. Pradisi, M. Nicolosi, and A. Sagnotti, ``On tadpoles and vacuum
redefinitions in string theory'', Nucl. Phys. B708 (2005) 3-44, \htmladdnormallink{arXiv:hep-th/0410101}{http://arxiv.org/abs/hep-th/0410101}

\bibitem{PST}P. Pasti, D. Sorokin, and M. Tonin, "On Lorentz invariant actions for chiral
p-forms", Phys. Rev. D55 (1997) 6292-6298, \htmladdnormallink{arXiv:hep-th/9611100}{http://arxiv.org/abs/hep-th/9611100}

\bibitem{WessBagger}J. Wess and J. Bagger, "Supersymmetry And Supergravity", Princeton University
Press, Princeton, 1992

\bibitem{Bilal}A. Bilal, ``Introduction to Supersymmetry'', \htmladdnormallink{arXiv:hep-th/0101055v1}{http://arxiv.org/abs/hep-th/0101055v1}

\bibitem{CremmerSuperYM}E. Cremmer, S. Ferrara, L. Girardello and A. Van Proeyen, "Yang-Mills Theories
With Local Supersymmetry: Lagrangian, Transformation Laws And Superhiggs Effect".
Nucl. Phys. B 212 (1983) 413


\bibitem{Lust:2003ky}
  D.~L{\"u}st and S.~Stieberger,
  ``Gauge threshold corrections in intersecting brane world models,''
  Fortsch.\ Phys.\  {\bf 55} (2007) 427
  [arXiv:hep-th/0302221].




\bibitem{Gmeiner:2009fb}
  F.~Gmeiner and G.~Honecker,
  ``Complete Gauge Threshold Corrections for Intersecting Fractional D6-Branes:
  The Z6 and Z6' Standard Models,''
  Nucl.\ Phys.\  B {\bf 829}, 225 (2010)
  [arXiv:0910.0843 [hep-th]].


\bibitem{Dimopoulos}A. Arvanitaki, N. Craig, S. Dimopoulos, S. Dubovsky and J. March-Russell, ``String Photini at the LHC'', \htmladdnormallink{arXiv:hep-ph/0909.5440v1}{http://arxiv.org/abs/0909.5440v1}

\bibitem{magMixing}F. Br\"ummer, J. Jaeckel and V. V. Khoze, ``Magnetic Mixing: Electric Minicharges from Magnetic Monopoles'', \htmladdnormallink{arXiv:hep-ph/0905.0633v1}{http://arxiv.org/abs/0905.0633v1}



\bibitem{AbelGoodsell_U1s}S. A. Abel, M. D. Goodsell, J. Jaeckel, V. V. Khoze, A. Ringwald, ``Kinetic Mixing of the Photon with Hidden U(1)s in String Phenomenology'', \htmladdnormallink{arXiv:hep-ph/0803.1449v1}{http://arxiv.org/abs/0803.1449v1}


\bibitem{Goodsell:2010ie}
  M.~Goodsell, A.~Ringwald,
  ``Light hidden-sector U(1)s in string compactifications,''
  Fortsch.\ Phys.\  {\bf 58 } (2010)  716-720,
 \htmladdnormallink{arXiv:1002.1840 [hep-th]}{http://arxiv.org/abs/1002.1840}



\bibitem{Williams:2011qb}
  M.~Williams, C.~P.~Burgess, A.~Maharana, F.~Quevedo,
  ``New Constraints (and Motivations) for Abelian Gauge Bosons in the MeV-TeV Mass Range,''
  \htmladdnormallink{arXiv:1103.4556 [hep-ph]}{http://arxiv.org/abs/1103.4556}
 


  


\bibitem{Alim:2009bx}
  M.~Alim, M.~Hecht, H.~Jockers, P.~Mayr, A.~Mertens, M.~Soroush,
  ``Hints for Off-Shell Mirror Symmetry in type II/F-theory Compactifications,''
  Nucl.\ Phys.\  {\bf B841 } (2010)  303-338, \htmladdnormallink{arXiv:0909.1842 [hep-th]}{http://arxiv.org/abs/0909.1842}

 



\bibitem{Grimm:2009ef}
  T.~W.~Grimm, T.~-W.~Ha, A.~Klemm, D.~Klevers,
  ``Computing Brane and Flux Superpotentials in F-theory Compactifications,"
  JHEP {\bf 1004 } (2010)  015, \htmladdnormallink{arXiv:0909.2025 [hep-th]}{http://arxiv.org/abs/0909.2025}


\bibitem{Grimm:2009sy}
  T.~W.~Grimm, T.~-W.~Ha, A.~Klemm, D.~Klevers,
  ``Five-Brane Superpotentials and Heterotic / F-theory Duality,"
  Nucl.\ Phys.\  {\bf B838 } (2010)  458-491,
  \htmladdnormallink{arXiv:0912.3250 [hep-th]}{http://arxiv.org/abs/0912.3250}
  
  




\bibitem{Jockers:2009ti}
  H.~Jockers, P.~Mayr, J.~Walcher,
  ``On N=1 4d Effective Couplings for F-theory and Heterotic Vacua,"
  \htmladdnormallink{arXiv:0912.3265 [hep-th]}{http://arxiv.org/abs/0912.3265}






\bibitem{Alim:2010za}
  M.~Alim, M.~Hecht, H.~Jockers, P.~Mayr, A.~Mertens, M.~Soroush,
  ``Type II/F-theory Superpotentials with Several Deformations and N=1 Mirror Symmetry,''
   \htmladdnormallink{arXiv:1010.0977 [hep-th]}{http://arxiv.org/abs/1010.0977}
   


\bibitem{Grimm:2010gk}
  T.~W.~Grimm, A.~Klemm, D.~Klevers,
  ``Five-Brane Superpotentials, Blow-Up Geometries and SU(3) Structure Manifolds,''
   \htmladdnormallink{arXiv:1011.6375 [hep-th]}{http://arxiv.org/abs/1011.6375}



\bibitem{Cremades:2002te}
  D.~Cremades, L.~E.~Ibanez and F.~Marchesano,
  ``SUSY quivers, intersecting branes and the modest hierarchy problem,''
  JHEP {\bf 0207} (2002) 009, \htmladdnormallink{arXiv:hep-th/0201205}{http://arxiv.org/abs/hep-th/0201205}.
 



\bibitem{KatzKachruSuperpot}S. Kachru, S. Katz, ``Open string instantons and superpotentials'', 
\htmladdnormallink{arXiv:hep-th/9912151v2}{http://arxiv.org/abs/hep-th/9912151v2}

\bibitem{SuperPotandMirrorSymm}B. Acharya, M. Aganagic, K. Hori and C. Vafa, ``Orientifolds, Mirror Symmetry
and Superpotentials'', \htmladdnormallink{arXiv:hep-th/0202208v1}{http://arxiv.org/abs/hep-th/0202208v1}

\bibitem{DbranesOnQuintic}I. Brunner, M.R. Douglas, A. Lawrence and C. R\"omelsberger, "D-branes on the Quintic", \htmladdnormallink{arXiv:hep-th/9906200}{http://arxiv.org/abs/hep-th/9906200}

\bibitem{KachruKatzMcGreevy_MirrorSymm}S. Kachru, S. Katz, A. E. Lawrence and J. McGreevy, "Open string instantons and superpotentials", Phys. Rev. D 62, 026001 (2000) \htmladdnormallink{arXiv:hep-th/9912151}{http://arxiv.org/abs/hep-th/9912151};
"Mirror symmetry for open strings", Phys. Rev. D 62 (2000) 126005 \htmladdnormallink{arXiv:hep-th/0006047}{http://arxiv.org/abs/hep-th/0006047}


\bibitem{Martucci}L. Martucci, ``D-branes on general $\mathcal{N}=1$ backgrounds: superpotentials and D-terms'', \htmladdnormallink{arXiv:hep-th/0602129}{http://arxiv.org/abs/hep-th/0602129v3}

\bibitem{Thomas}R. P. Thomas, "Moment maps, monodromy and mirror manifolds",
\htmladdnormallink{arXiv:math.dg/0104196}{http://arxiv.org/abs/math.dg/0104196}


\bibitem{HaackLuest}M. Haack, D. Krefl, D. L\"ust, A. Van Proeyen and
M. Zagermann, ``Gaugino condensates and D-terms from D7-branes'', \htmladdnormallink{JHEP 01 (2007) 078}{http://iopscience.iop.org/1126-6708/2007/01/078}

\bibitem{Villadoro}G. Villadoro and F. Zwirner, ``D terms from D-branes, gauge invariance and moduli
stabilization in flux compactifications'', JHEP 03 (2006) 087 \htmladdnormallink{arXiv:hep-th/0602120}{http://arxiv.org/abs/hep-th/0602120};\\
G. Villadoro and F. Zwirner, ``N = 1 effective potential from dual type-IIA D6/O6 orientifolds
with general fluxes'', JHEP 0506 (2005) 047, \htmladdnormallink{arXiv:hep-th/0503169}{http://arxiv.org/abs/hep-th/0503169}





\bibitem{Grimm:2011dx}
  T.~W.~Grimm and D.~V.~Lopes,
  ``The N=1 effective actions of D-branes in Type IIA and IIB orientifolds,''
  arXiv:1104.2328 [hep-th].



\bibitem{FerraraSymplStructN=2}A. Ceresole, R. D'Auria and S. Ferrara, "The Symplectic Structure of N=2 Supergravity and its Central Extension", \htmladdnormallink{arXiv:hep-th/9509160v1}{http://arxiv.org/abs/hep-th/9509160v1}

\bibitem{FerraraDualitySugraYM}A.Ceresole, R.D'Auria, S.Ferrara and A.Van Proeyen, ``Duality transformations in supersymmetric Yang-Mills theories coupled to supergravity'', Nucl. Phys. B444 (1995) 92

\bibitem{VanProeyenSpecialKaehler}B. Craps, F. Roose, W. Troost and A. Van Proeyen, "What is special Kaehler
geometry?", Nucl. Phys. B 503 (1997) 565 \htmladdnormallink{arXiv:hep-th/9703082}{http://arxiv.org/abs/hep-th/9703082}

\bibitem{FreedSpecialKaehler}D. S. Freed, "Special Kaehler manifolds," Commun. Math. Phys. 203, 31 (1999)
\htmladdnormallink{arXiv:hep-th/9712042}{http://arxiv.org/abs/hep-th/9712042}


\bibitem{Suzuki}H. Suzuki, "Calabi-Yau Compactification of
Type IIB String and a Mass Formula for the
Extreme Black Hole", preprint OU-HET-220,
\htmladdnormallink{arXiv:hep-th/9508001}{http://arxiv.org/abs/hep-th/9508001}






\end{thebibliography}
\end{document}